\documentstyle[12pt]{article}
\newcommand{\fracd}[2]{\frac{\displaystyle {#1}}{\displaystyle {#2}}}
\newcommand{\E}{\mbox{\rm E}}
\newcommand{\sinc}{\mbox{\rm sinc}}
\newcommand{\sgn}{\mbox{\rm sgn}}
\begin{document}
%\begin{flushleft}
%\noindent
%To appear in {\it Fractals} (1997)
%\end{flushleft}

\begin{center}
{\Large \bf ANALYSIS, SYNTHESIS, AND ESTIMATION OF FRACTAL-RATE
STOCHASTIC POINT PROCESSES
%\footnote{submitted to Fractals}
}
\end{center}

\begin{center}
{\large
STEFAN THURNER, STEVEN B. LOWEN, MARKUS C. FEURSTEIN\\
and CONOR HENEGHAN}\\
{\it Department of Electrical \& Computer Engineering,
Boston University, Boston, MA 02215}\\[12pt]
{\large
HANS G. FEICHTINGER}\\
{\it Institut f\"ur Mathematik,
Universit\"at Wien,
A-1090 Vienna, Austria}\\[12pt]
{\large
MALVIN C. TEICH
\footnote{Corresponding author. Email address: teich@bu.edu;\\
URL: http://ece.bu.edu/ECE/faculty/homepages/teich.html/}
}\\
{\it Departments of Electrical \& Computer Engineering,
Biomedical Engineering, and Physics,
Boston University, Boston, MA 02215}\\[12pt]
\end{center}

\section*{Abstract}

Fractal and fractal-rate stochastic point processes (FSPPs and
FRSPPs) provide useful models for describing a broad range of diverse
phenomena, including electron transport in amorphous semiconductors,
computer-network traffic, and sequences of neuronal action potentials.
A particularly useful statistic of these processes is the fractal
exponent $\alpha$, which may be estimated for any FSPP or FRSPP by
using a variety of statistical methods.
Simulated FSPPs and FRSPPs consistently exhibit bias in this fractal
exponent, however, rendering the study and analysis of these
processes non-trivial.
In this paper, we examine the synthesis and estimation of FRSPPs by
carrying out a systematic series of simulations for several different
types of FRSPP over a range of design values for $\alpha$.
The discrepancy between the desired and achieved values of $\alpha$
is shown to arise from finite data size and from the character of
the point-process generation mechanism.
In the context of point-process simulation, reduction of this
discrepancy requires generating data sets with either a large number
of points, or with low jitter in the generation of the points.
In the context of fractal data analysis, the results presented here
suggest caution when interpreting fractal exponents estimated from
experimental data sets.
%PACS number(s): 05.40.+j, 02.50.-r, 87.10.+e, 72.80.-r
\pagebreak

\section{INTRODUCTION}

Some random phenomena occur at discrete times or locations, with the
individual events largely identical, such as a sequence of neural
action potentials.
A stochastic point process \cite{COX80} is a mathematical
construction which represents these events as random points in a
space.
Fractal stochastic point processes exhibit scaling in all of
the statistics considered in this paper; fractal-rate stochastic
point processes do so in some of them.
In this work we consider the simulation and estimation of fractal
and fractal-rate stochastic point processes on a line, which model a
variety of observed phenomena in the physical and biological sciences.
This work provides an extension and generalization of an earlier
paper along these lines \cite{LOW95}.

\subsection{Mathematical Descriptions of Stochastic Point
Processes}
\label{zkdef}

Figure~\ref{basics} shows several representations that are useful in
the analysis of point processes.
Figure~\ref{basics}(a) demonstrates a sample function of a point
process as a series of impulses occurring at specified times $t_n$.
Since these impulses have vanishing width, they are most rigorously
defined as the derivative of a well-defined counting process $N(t)$
[Fig.~\ref{basics}(b)], a monotonically increasing function of $t$,
that augments by unity when an event occurs.
Accordingly, the point process itself is properly written as $dN(t)$,
since it is only strictly defined within the context of an integral.

The point process is completely described by the set of event times
$\{ t_n \}$, or equivalently by the set of interevent intervals.
However, the sequence of counts depicted in Fig.~\ref{basics}(c)
also contains much information about the process.
Here the time axis is divided into equally spaced contiguous counting
windows of duration $T$ sec to produce a sequence of counts
$\{ Z_{k} \}$, where $Z_{k} = N[(k+1)T] - N[kT]$ denotes the number of
events in the $k$th window.
As illustrated in Fig.~\ref{basics}(d), this sequence forms a
discrete-time random process of nonnegative integers.
In general, information is lost in forming the sequence of counts,
although for a regular point process the amount lost can be made
arbitrarily small by reducing the size of the counting window $T$.
An attractive feature of this representation is that it preserves the
correspondence between the discrete time axis of the counting process
$\{ Z_{k} \}$ and the absolute ``real'' time axis of the underlying
point process.
Within the process of counts $\{ Z_{k} \}$, the elements $Z_{k}$ and
$Z_{k+n}$ refer to the number of counts in windows separated by
precisely $T(n-1)$ sec, so that correlation in the process
$\{ Z_{k} \}$ is readily associated with correlation in the
underlying point process $dN(t)$.

\subsection{Scaling and Power-Law Behavior in Fractal Stochastic Processes}
\label{exhscaling}
The complete characterization of a stochastic process involves a
description of all possible joint probabilities of the various events
occurring in the process.
Different statistics provide complementary views of the process; no
single statistic can in general describe a stochastic process
completely.
Fractal stochastic processes exhibit scaling in their statistics.
Such scaling leads naturally to power-law behavior, as demonstrated
in the following.
Consider a statistic $f$ which depends continuously on the scale $x$
over which measurements are taken.
Suppose changing the scale by any factor $a$ effectively scales the
statistic by some other factor $g(a)$, related to the factor but
independent of the original scale:
\begin{equation}
f(ax) = g(a)f(x).
\end{equation}
The only nontrivial solution of this scaling equation for real
functions and arguments, that is independent of $a$ and $x$ is
\begin{equation}
f(x) = bg(x) \hspace{7mm} \mbox{with} \hspace{7mm} g(x) = x^{c}
\label{scale0}
\end{equation}
for some constants $b$ and $c$ \cite{LOW95,RUD76}.
Thus statistics with power-law forms are closely related with the
concept of a fractal. The particular case of fixed $a$ admits a more
general solution \cite{SHLE91}:
\begin{equation}
g(x;a)= x^{c} \, {\rm cos}[2\pi \, {\rm ln}(x)/{\rm ln}(a)] \, .
\end{equation}

\subsection{Fractal Stochastic Point Processes (FSPPs) }

Consider, for example, a commonly encountered first-order statistic
for a stochastic point process, the interevent interval histogram
(IIH).
This estimates the interevent-interval probability density function
(IIPDF) $p(t)$ by computing the relative frequency of occurrence of
interevent intervals as a function of interval size.
This measure highlights the behavior of the times between adjacent
events, but reveals none of the information contained in the
relationships among these times, such as correlation between adjacent
time intervals.
For a fractal stochastic point process (FSPP) the IIPDF follows the
form of Eq.~(\ref{scale0}), so that $p(t) \sim t^{c}$ over a certain
range of $t$, where $c< -1$.

A number of statistics may be used to describe an FSPP, and each
statistic which scales will in general have a different scaling
exponent $c$.
Each of these exponents can be simply related to a more general
parameter $\alpha$, the fractal exponent, where the
exact relation between these two exponents will depend upon the
statistic in question.
For example, the exponent $c$ of the IIPDF defined above is related
to the fractal exponent $\alpha$ by $c = -(1 + \alpha)$.

Sample functions of the fractal renewal point process (FRP; see
Sec.~\ref{frp_def}) are true fractals;
the expected value of their generalized dimensions
(see Sec.~\ref{othermeas}) assumes a nonintegral value between the
topological dimension (zero) and the Euclidean dimension (unity)
\cite{MAN83}.

\subsection{Fractal-Rate Stochastic Point Processes (FRSPPs)}
\label{frsppdef}
However, the sequence of unitary events observed in many biological
and physical systems do not exhibit power-law-distributed IIHs but
nevertheless exhibit scaling in other statistics.
These processes therefore have integral generalized dimensions (see
Sec.~\ref{othermeas}) and are consequently not true fractals.
They are nevertheless endowed with rate functions that are indeed
either fractals or their increments: fractal Gaussian noise, fractal
Brownian motion, or closely related processes.
Therefore, such point processes are more properly termed
fractal-{\it rate} stochastic point processes (FRSPPs).

\subsection{Relation of Fractal Exponent to Hurst Exponent}

The fractal exponent $\alpha$ defined above is related to the more
commonly encountered Hurst exponent $H$ \cite{HUR51}.
The relationship is ambiguous, however, since some authors
\cite{MAN83,FLA89,FLA92A,FLA92B,WOR92} use the formula
$\alpha = 2H + 1$ for all values of $\alpha$, while others
\cite{BAR88} use $\alpha = 2H - 1$ for $\alpha < 1$ to restrict
$H$ to the range $(0,1)$.
In this paper, we avoid this confusion by considering $\alpha$
directly instead of $H$.

\section{APPLICATIONS OF FRACTAL AND FRACTAL-RATE STOCHASTIC POINT PROCESSES}

Many phenomena are readily represented by FSPPs and FRSPPs.
We provide several examples drawn from the physical
\cite{PRE78,BUC83,WEI88} and biological \cite{BAS94,WES94,AND96}
sciences.

\subsection{Trapping Times in Amorphous Semiconductors}

A multiple trapping model has been used to show how traps that are
exponentially distributed over a large range of energies lead to a
power-law decay of current in an amorphous semiconductor
\cite{PFI78,ORE82,KAS85,TIE80,SHL87}.
If a pulse of light strikes such a semiconductor, the many carriers
excited out of their traps will be available to carry current until
they are recaptured, which happens relatively quickly.
At some point each carrier will be released from its trap by thermal
excitation and become mobile for a time, and then be recaptured by
another trap.
For exponentially distributed energy states with identical capture
cross sections, the electrons tend to be trapped in shallow states at
first, but the probability of being caught in a deep trap increases
as time progresses.
This leads to a current that decreases as a power-law function of
time.

The multiple trapping model may be usefully recast in terms of a
standard fractal renewal process (see Sec.~\ref{frp_def})
\cite{LOW93A,LOW92A}.
Consider an amorphous semiconductor with localized states (traps)
with energies which
are exponentially distributed with parameter $E_0$ between a
minimum energy $E_L$ of the order of $\kappa {\cal T}$, where
$\kappa$ is Boltzmann's constant and ${\cal T}$ is the temperature in
degrees Kelvin; and a maximum energy $E_H$ determined by the bandgap
of the material.
For a particular trap with energy ${\cal E}$, the corresponding mean
waiting time to release is
\begin{equation}
\tau = \tau_0 \exp({\cal E} / \kappa {\cal T}),
\end{equation}
where $\tau_0$ is the average vibrational period of the atoms in the
semiconductor.
If we define characteristic time cutoffs
$A \equiv \tau_0 \exp(E_L/ \kappa {\cal T})$ and
$B \equiv \tau_0 \exp(E_H/ \kappa {\cal T})$, and the power-law exponent
$c \equiv \kappa {\cal T} / E_0$, then the mean waiting time $\tau$ has a
density which decays as a power law with this exponent between those
two cutoffs.
Each trap holds carriers for times that are exponentially distributed
given the conditional parameter $\tau$, and averaging this
exponential density over all possible values of $\tau$ yields the
unconditional trapping time density, which is itself approximately
power law:
\begin{equation}
p(t) \approx c \Gamma(c+1)A^{c} t^{-(c+1)},
\end{equation}
for $A \ll t \ll B$.
Thus each carrier will be trapped for a period that is essentially
power-law distributed.

Upon escaping from a trap, the carrier can conduct current for a
short time until it is again captured by another trap.
Thus each carrier executes a series of current-carrying jumps well
described by a standard FRP.
Assuming that each carrier acts independently of the others, the
action of the carriers as a whole can be modeled as the superposition
of a collection of such component processes, which converges to the
fractal-Gaussian-noise driven Poisson process (FGNDP; see
Secs.~\ref{fgn} and \ref{fdspp}) in the limit of a large number of
carriers.
Again, both experimental \cite{BHA90} and theoretical
\cite{SHL87,TOM90} results point to a power-law or fractal decay in
the power spectral density, while several other statistics also show
scaling behavior.

\subsection{Noise and Traffic in Communications Systems}

Burst noise occurs in many communications systems and is characterized
by relatively brief noise events which cluster together, separated by
relatively longer periods of quiet.
Berger and Mandelbrot \cite{BER63,MAN65} long ago showed that burst
errors in communication systems are well modeled by a version of a
fractal renewal process (see Sec.~\ref{frp_def}), and in particular
that the interevent times were essentially independent of each other
for time scales determined by the resolution and the duration of the
observation.

Moreover, the rate of traffic flow itself displays fractal
fluctuations on a variety of high-speed packet-switching networks
conducting different types of traffic
\cite{LEL93,ERR93,RYU94,LEL94,BER95}.
This has been demonstrated for time scales greater than about
one sec in both the power spectral density and the Fano factor
(see Secs.~\ref{psdest} and~\ref{ffest}).
Over these time scales, the fractal-shot-noise driven Poisson
process (FSNDP; see Secs.~\ref{fsn} and \ref{fdspp})
has been successfully employed to
model the traffic \cite{RYU95}.

\subsection{Current Flow in Biological Ion Channels}
\label{ionch}
Ion channels reside in biological cell membranes, permitting ions to
diffuse in or out \cite{SAK83}.
These channels are usually specific to a particular ion, or group of
related ions, and block the passage of other kinds of ions.
Further, most channels have gates, and thus the channels may be
either open or closed.
In many instances, intermediate conduction states are not observed.
Some ion channels may be modeled by a two-state Markov process
\cite{DEF93}, with one state representing the open channel, and the
other representing the closed channel.
This model generates exponentially distributed dwell times in both
states, which are, in fact, sometimes observed.
However, many ion channels exhibit independent power-law distributed
closed times between open times of negligible duration
\cite{LAU88,MIL88,LIE90,TEI89}, and are well described by a fractal
renewal point process (see Sec.~\ref{frp_def}) \cite{LOW93B,LOW93C}.

\subsection{Vesicular Exocytosis at the Synaptic Cleft}
\label{exocytosis}
Communication in the nervous system is usually mediated by the
exocytosis of multiple vesicular packets (quanta) of neurotransmitter
molecules at the synapse between two cells, triggered by an action
potential (information-bearing signal) at the presynaptic cell
\cite{KAT66}.
Even in the absence of an action potential, however, many neurons
spontaneously release individual packets of neurotransmitter
\cite{FAT52}.
On arrival at the postsynaptic membrane, each neurotransmitter packet
induces a miniature endplate current (MEPC); superpositions of
MEPC-like events comprise the postsynaptic endplate currents elicited
by nerve impulses arriving at the presynaptic cell \cite{DEL54}.
Analysis of the sequence of MEPC and MEPC-like events in a variety of
preparations reveals the presence of memory over a range of
times and frequencies \cite{LOW97A}.
The fractal-lognormal-noise driven Poisson process model (FLNDP;
see Secs.~\ref{fln} and \ref{fdspp}) provides an excellent model for
this process, for a variety of statistical measures
\cite{LOW97A,LOW97B}.

\subsection{Action Potentials in Isolated Neuronal Preparations}
Many biological neurons carry information in the form of sequences of
action potentials, which are localized regions of depolarization
traveling down the length of an axon.
Action potentials or neural spikes are brief and largely identical
events and are well represented by a point process.
The firing patterns of an isolated neuron and an isolated axons have
been shown to have fractal characteristics.
Musha and colleagues generated a synthetic spike train consisting of
electrical impulses separated by independent intervals drawn from the
same Gaussian distribution \cite{MUS81}.
This nonfractal stimulus as applied to one end of an excised squid
giant axon; the membrane voltage at the far end was considered to be
the response spike train.
The resulting power spectral density followed a $1/f$-type decay.
Spontaneous firing from an excised giant snail neuron yielded similar
results \cite{MUS83}, although for this preparation the data was
selected to reduce fluctuations, and analyzed by interevent number
rather than in real time.

\subsection{Auditory-Nerve-Fiber Action Potentials}
More recently, fractal behavior has been observed in the sequence of
action potentials recorded from several {\it in vivo} vertebrate
preparations.
Real-time recordings were made under both spontaneous and driven
firing conditions.
Over short time scales, nonfractal stochastic point processes prove
adequate for representing such nerve spikes, but over long time
scales (typically greater than one sec) the fractal nature becomes
evident \cite{LOW92B}.
Estimators of the rate of the process converge more slowly than for
nonfractal processes, displaying fluctuations which decrease as a
power-law function of the time used to estimate the rate
\cite{TEI90B}.
With the inclusion of the refractory effects of nerve fibers, FRSPP
models have been shown to properly characterize the statistical
properties of certain potentials in peripheral primary auditory fibers in
several species, over all time scales and for a broad variety of
statistical measures
\cite{TEI89,TEI90B,TEI90A,POW91,POW92,TEI92,KUM93,TEI94,KEL94,KEL96,LOW96B,LOW96D,LOW96A};
only four parameters are required.
This process possibly arises from superpositions of fractal
ion-channel transitions \cite{LOW93B,LOW93C} or by fractal vesicular
exocytosis in inner-ear sensory cells, as described briefly in
Secs.~\ref{ionch} and ~\ref{exocytosis}, respectively.

\subsection{Optic-Nerve-Fiber Action Potentials}
Many neurons in the mammalian visual system transmit information by
means of action potentials as well, and FRSPPs also provide suitable
models for describing the behavior of these neurons \cite{TEI96B}.
The gamma renewal process, which is nonfractal, has proven to be a
useful model for some of these processes over short time scales
\cite{TRO92}.
However, nerve spike trains recorded from both cat
retinal-ganglion-cell and lateral-geniculate-nucleus neurons, like
those recorded from primary auditory neurons, exhibit fractal
behavior over time scales greater than about $1$ sec, thus
necessitating the use of an FRSPP description \cite{TEI96B}.
Long-duration temporal correlation has also been observed in spike
trains from cat striate-cortex neurons \cite{TEI95} and from an
insect visual interneuron \cite{TUR95}.

\subsection{Central-Nervous-System Action Potentials}
\label{cns}
Action-potential data collected from rabbit somatosensory cortical
neurons and cat respiratory control neurons show experimental IIH
plots which decay as power laws \cite{WIS81}.
In addition, for the data lengths considered, serial correlation
coefficients among interevent intervals from these data sets do not
differ from zero in a statistically significant way.
(However, differences may well emerge for longer data sets, as has
been seen, for example, in auditory-nerve-fiber recordings
\cite{LOW92B}.)
These two interevent-interval characteristics --- independence and a
power-law distribution --- define the fractal renewal point process
(see Sec.~\ref{frp_def}), which is an FSPP.

On the other hand, action potentials in the cat mesencephalic
reticular formation (MRF) exhibit activity that can be modeled by an
FRSPP.
During REM phases of sleep, these neurons exhibit $1/f$-type spectra
\cite{YAM86,GRU89,GRU93}, as do certain hippocampal and thalamic
neurons \cite{KOD89}.
The cluster Poisson point process (see Sec.~\ref{cluster}) has been
used successfully to model MRF neural activity \cite{GRU93}.
Furthermore, there is some evidence that these neurons have IIHs
whose tails decay in power-law fashion \cite{YAM83}, so that a type
of FSPP may prove suitable for describing these spike trains.

\subsection{Human Heartbeat Times}
\label{heartbeat}
The sequence of human heartbeats exhibits considerable variability
over time, both in the short-term and in the long-term
patterns of the times between beats \cite{KOB82,BER86}.
A point process of the heartbeats is formed by focusing on the times
between maximum contractions (R-R intervals).
A particular FRSPP with an integrate-and-fire
substrate has been constructed and shown to successfully
describe these events \cite{TUR93,TUR96}.
As with auditory and visual-system spike trains,
over short time scales nonfractal point processes provide suitable
models for the pattern of times between contractions; for times
longer than roughly $10$ sec, only fractal models suffice.
Interestingly, parameters of the FRSPP used to model the data
indicate promise for the diagnosis of various disease states of the
heart \cite{TUR96}.

\section{ANALYSIS OF FRACTAL-RATE STOCHASTIC\newline
POINT PROCESSES}
\label{analysis}
As the examples described above illustrate, many phenomena are
amenable to modeling by FRSPPs.
The value of the fractal exponent $\alpha$ can often provide
important information regarding the nature of an underlying process,
and can also serve as a useful classification tool in some cases, as
with human heartbeat data as mentioned in Section~\ref{heartbeat}.
Accordingly, it is desirable to estimate $\alpha$ reliably,
although this task is often confounded by a variety
of issues \cite{LOW95,SCH92,BER92}.

Here we briefly review some of the techniques used for estimating
$\alpha$.
By way of illustration, we apply these methods to a train of action
potentials recorded from a lateral-geniculate-nucleus (LGN) relay
neuron in the cat visual system.
There are 24285 events in this particular spike train, with an
average interevent interval of 0.133 sec, comprising a total duration
of 3225 sec \cite{TEI96B}.

\subsection{Interval Probability Density Function}
The interevent-interval histogram (IIH), an estimate of the interval
probability function, and the generalized dimension (GD)
\cite{HEN83,GRA83,THE90}, are useful measures for fractal stochastic
point processes, and in particular for the fractal renewal process
(see Sec.~\ref{frp_def}).
The IIH is simply the relative frequency of interevent-interval
occurrence in the data set, ignoring all correlations among the
intervals.

Since realizations of FSPPs form true fractals, all measures
considered in this paper, including the IIH, exhibit scaling for
these processes.
Thus the IIH proves useful in elucidating the fractal nature of FSPPs.
For FRSPPs, on the other hand, the fractal structure is not
manifested in first-order interevent-time statistics, so that the IIH
does not scale over a significant range of interevent times.
Rather, the fractal character of these processes lie in the
dependencies among the interevent intervals, which are captured by the
measures considered in Secs.~\ref{crsec}--\ref{afest}.

\subsection{Generalized Dimension}
\label{othermeas}
The generalized dimension is of interest for fractal stochastic point
processes.
If a data segment of length $L$ is divided into
intervals of length $T$, with $Z_k$ representing the number of points
in the $k$th interval (see Fig.~1(c)), then the generalized dimension
$D_q$ of a point process is defined as
\begin{equation}
D_q \equiv \frac{1}{q-1} \lim_{T \to 0}
\frac{\log\left(\sum Z_k^q\right)}{\log(T)},
\end{equation}
where the sum extends over all non-empty intervals.
Particular cases are the capacity or box-counting dimension $D_0$,
the information dimension $\lim_{q \to 1} D_q$, and the correlation
dimension $D_2$.
The sum is a form of time averaging; for a stochastic point process,
it is convenient to replace the sum by the product of $L/T$ and the
expected value of $Z^q$.
For a fractal-rate point process, however, analytical values of the
$D_q$ will not in general equal analytical values of the dimensions
obtained from the PG or AF, and in fact the $D_q$ will often assume
integer values (see Sec.~\ref{frsppdef}).

\subsection{Coincidence Rate}
\label{crsec}
The CR measures the correlations between pairs of events with a
specified time delay between them, regardless of intervening events,
and is related to the autocorrelation function used with continuous
processes.
The CR is defined as
\begin{equation}
G(\tau) \equiv \lim_{\Delta \rightarrow 0} \frac{\mbox{Pr} \left\{
{\cal E}[0,\Delta] \mbox{ and } {\cal E}[\tau,\tau+\Delta] \right\}}
{\Delta^{2}},
\end{equation}
where ${\cal E}[s,t]$ denotes the occurrence of at least one event of
the point process in the interval $[s,t)$ and $\tau$ is the delay time.
For an ideal fractal or fractal-rate stochastic point process with a
fractal exponent $\alpha$ in the range $0<\alpha<1$, the coincidence
rate assumes the form
\cite{LOW95}
\begin{equation}
G(\tau) = \lambda \delta(\tau) + \lambda^{2}
\left[1 + (\mid \tau\mid /\tau_{0})^{\alpha-1}\right],
\label{fractalcr}
\end{equation}
where $\lambda$ is the mean rate of the process, $\delta(\tau)$
denotes the Dirac delta function, and $\tau_{0}$ is a constant
representing the fractal onset time.

A stationary, regular point process with a CR following this form for
{\it all} delay times $\tau$ exhibits infinite mean.
Further, statistics of fractal data sets collected from experiments
exhibit scaling only over a finite range of times and frequencies,
as determined by the resolution of the experimental apparatus
and the duration of the experiment.
Nevertheless, in much of the following we employ
Eq.~(\ref{fractalcr}) without cutoffs since we find that the cutoffs
do not significantly affect the mathematical results in many cases.
We employ similar ideal forms for other second-order measures defined
later in this paper for the same reasons.

The coincidence rate can be directly estimated from its definition.
However, in practice the CR is a noisy measure, since its definition
essentially involves a double derivative.
Furthermore, for FSPPs and FRSPPs typical of physical and biological
systems, the CR exceeds its asymptotic value $\lambda^{2}$ by only a
small fraction at any large but practical value of $\tau$, so that
determining the fractal exponent with this small excess presents
serious difficulties.
Therefore we do not specifically apply this measure to the LGN data,
although the formal definition of coincidence rate plays a useful
role in developing other, more reliable measures.

\subsection{Power Spectral Density}
\label{psdest}
The power spectral density (PSD) is a familiar and well-established
measure for continuous-time processes.
For point processes the PSD and the CR introduced above form a
Fourier transform pair, much like the PSD and the autocorrelation
function do for continuous-time processes.
The PSD provides a measure of how the power in a process is
concentrated in various frequency bands.
For an ideal fractal point process, the PSD assumes the form
\begin{equation}
S(\omega) = \lambda^2 \delta(\omega/2\pi)
+ \lambda \left[1+(\omega/\omega_0)^{-\alpha}\right],
\label{fractalpsd}
\end{equation}
for relevant time and frequency ranges, where $\lambda$ is the mean
rate of events and $\omega_0$ is a cutoff radian frequency.

The PSD of a point process can be estimated using the periodogram (PG)
$S_{Z}(\omega)$ of the sequence of counts, rather than from the point
process itself \cite{MAT82}.
This method introduces a bias at higher frequencies, since the fine
time resolution information is lost as a result of the minimum
count-window size.
Nevertheless, since estimating the fractal exponent principally
involves lower frequencies where this bias is negligible, and
employing the sequence of counts permits the use of vastly more
efficient fast Fourier transform methods, we prefer this technique.
Alternate definitions of the PSD for point processes (and thus for
the PG used to estimate them) exist; for example, a different PSD may
be obtained from the real-valued discrete-time sequence of the
interevent intervals.
However, features in this PSD cannot be interpreted in terms of
temporal frequency \cite{TUR96}.

Figure~\ref{vispp} displays the PG for the visual system LGN data
calculated using the count-based approach.
Throughout the text of this paper we employ radian frequency
$\omega$ (radians per unit time) to simplify the analysis, while
figures are plotted in common frequency $f=\omega/2\pi$ (cycles per
unit time) in accordance with common usage.
For low frequencies, the PG decays as $1/\omega^{\alpha}$, as
expected for a fractal point process.
Fitting a straight line (shown as dotted) to the doubly logarithmic
plot of the PG, over the range from 0.002 Hz to 1 Hz, provides an
estimate $\widehat{\alpha} \approx 0.67$.

\subsection{Fano Factor}
\label{ffest}
Another measure of correlation over different time scales is provided
by the Fano factor (FF), which is the variance of the number of
events in a specified counting time $T$ divided by the mean number of
events in that counting time.
This measure is sometimes called the
index of dispersion of counts \cite{COX80}.
In terms of the sequence of counts illustrated in
Fig.~\ref{basics}(c), the Fano factor is simply the variance of
$\{ Z_{k} \}$ divided by the mean of $\{ Z_{k} \}$, i.e.,
\begin{equation}
F(T) \equiv \frac{\E[Z_{k}^{2}] -\E^{2}[Z_{k}]}{\E[Z_{k}]},
\end{equation}
where $E[\cdot]$ represents the expectation operation.
The FF may be obtained from the coincidence rate by an integral
transform
\begin{equation}
F(T) = 1 + \frac{2}{\lambda T} \int_{0+}^T (T - \tau)
\left[G(\tau) - \lambda^2\right]\, d\tau,
\end{equation}
where $\lambda$ is the average rate of events of the point process,
and the lower limit of the integral approaches zero from the positive
side \cite{COX80}.

The FF varies as a function of counting time $T$.
The exception is the homogeneous Poisson point process (HPP), which
is important as a benchmark in point-process theory just as the
Gaussian is in the theory of continuous stochastic processes.
For an HPP, the variance-to-mean ratio is always unity for any
counting time $T$.
Any deviation from unity in the value of $F(T)$ therefore indicates
that the point process in question is not Poisson in nature.
An excess above unity reveals that a sequence is less ordered than an
HPP, while values below unity signify sequences which are more
ordered.
For an ideal FSPP or FRSPP with $0 < \alpha < 1$, the FF assumes the
form
\begin{equation}
F(T) = 1 + (T/T_0)^{\alpha},
\label{fractalff}
\end{equation}
where $T_0$ is a fractal onset time that differs from $\tau_0$ in
Eq.~(\ref{fractalcr}) [see Eq.~(\ref{relati})].
Therefore a straight-line fit to an estimate of $F(T)$ vs.\ $T$ on a
doubly logarithmic plot can also be used to estimate the fractal
exponent.
However, the estimated slope of the FF saturates at unity, so that
this measure finds its principal applicability for processes with
$\alpha < 1$ \cite{LOW95,LOW96B,TEI96B,LOW93D}.

Figure~\ref{visrff} shows the estimated FF curve for the same data
set as shown in Fig.~\ref{vispp}.
For counting times $T$ greater than approximately 0.3 sec, this curve
behaves essentially as $\sim T^{\alpha}$.
The estimated value is $\widehat{\alpha} = 0.66$ (dotted line),
closely agreeing with the value obtained from the PG.
Fair agreement between these two measures exists in other sensory
spike-train and heartbeat sequence data sets
\cite{TEI92,TEI96B,TUR96} and in simulated FRSPPs \cite{LOW95}.

\subsection{Allan Factor}
\label{afest}
The Allan variance, as opposed to the ordinary variance, is defined
in terms of the variability of {\it successive} counts
\cite{ALL66,BAR66}.
In analogy with the Fano factor (FF), we define the Allan factor (AF)
for the sequence of counts shown in Fig.~\ref{basics} as
\begin{equation}
A(T) \equiv \frac{\E\left[(Z_{k+1}-Z_{k})^{2}\right]}{2\E[Z_{k}]}.
\label{afdef}
\end{equation}
In terms of quantities defined earlier, we have
\begin{equation}
A(T) = 2F(T) - F(2T)
= 1 + \frac{4}{\lambda T} \int_{0+}^T (T - \tau)
\left[G(\tau) - G(2\tau)\right]\, d\tau.
\label{afcalc}
\end{equation}
As for the FF, the value of the Allan factor for the HPP is unity.
For an ideal FRSPP with $0 < \alpha < 3$, the AF also assumes a
power-law-varying form
\begin{equation}
A(T) = 1 + (T/T_1)^{\alpha},
\label{fractalaf}
\end{equation}
where $T_1$ is another fractal onset time, different from $\tau_0$
and $T_0$ [see Eq.~(\ref{relati})];
therefore a straight line fit of an estimate of
$A(T)$ vs.\ $T$ on a doubly logarithmic plot yields yet another
estimate of the fractal exponent.

Figure~\ref{visaff} shows the estimated Allan factor curve for the
same data set as shown in Figs.~\ref{vispp} and~\ref{visrff}.
From visual inspection, this measure appears to be considerably
``rougher'' than the FF, which is typical of the data sets we have
analyzed.
Nevertheless it is clear that for counting times $T$ greater than
approximately 0.3 sec, its behavior can also be approximated as
$\sim T^{\alpha}$.
To estimate the value of $\alpha$, a straight line fit to the doubly
logarithmic plot of the estimate of $A(T)$ vs.\ counting time $T$ was
provided.
The value of $\widehat{\alpha} = 0.65$ obtained agrees well with the
values calculated using the PG and FF.
Excellent agreement between the AF and the PG exists in other
sensory spike-train and heartbeat sequence data sets
\cite{LOW96B,TEI96B,TUR96,TEI96A} and simulations \cite{TEI96B}.
Furthermore, the AF appears to yield superior estimates of $\alpha$
than the FF in all cases \cite{LOW96B,TEI96B,TUR96,TEI96A}.
In particular, instead of saturating at a power-law exponent of
unity, estimates of $\alpha$ obtained from the AF can range up to a
value of three \cite{LOW96B}.
Thus we do not employ the FF in the remaining sections of this paper.

The Allan variance, $\E\left[(Z_{k+1}-Z_{k})^{2}\right]$ may be
recast as the variance of the integral of the point process under
study multiplied by the following function:
\begin{equation}
\psi_{\rm Haar}(t) =
\left\{ \begin{array}{ll}
-1 & \mbox{ for $-T < t < 0$,}\\
+1 & \mbox{ for $0 < t < T$,}\\
0 & \mbox{ otherwise.}
\end{array} \right.
\label{haar}
\end{equation}
Equation~(\ref{haar}) defines a scaled wavelet function, specifically
the Haar wavelet.
This can be generalized to any admissible wavelet $\psi(t)$; when
suitably normalized, the result is a wavelet Allan factor (WAF)
\cite{TEI96A,HEN96}.
This generalization enables an increased range of fractal exponents
to be estimated, at the cost of reducing the range of times over
which the WAF varies as $~T^\alpha$.
In particular, for a particular wavelet with regularity (number of
vanishing moments) $R$, fractal exponents $\alpha < 2R + 1$ can be
reliably estimated \cite{TEI96A,HEN96}.
For the Haar basis, $R=1$, while all other wavelet bases have $R>1$.
Thus the WAF employing bases other than the Haar should prove useful
in connection with FRSPPs for which $\alpha \ge 3$; for processes
with $\alpha < 3$, however, the AF appears to be the best choice
\cite{TEI96A,HEN96}.

\subsection{Relationships Among the Measures}
For an ideal FSPP or FRSPP with $0 < \alpha < 1$, any one of
Eqs.~(\ref{fractalcr}), (\ref{fractalpsd}), (\ref{fractalff}), and
(\ref{fractalaf}) implies the other three, with \cite{LOW95,LOW96C}
\begin{eqnarray}
\lambda \tau_0^{1-\alpha} T_0^\alpha & = & \alpha(\alpha+1) / 2
\nonumber \\
\omega_0^\alpha T_0^\alpha & = & \cos(\pi \alpha/2) \Gamma(\alpha+2)
\label{scale2} \\
T_0^\alpha T_1^{-\alpha} & = & 2 - 2^\alpha.
\nonumber
\label{relati}
\end{eqnarray}
For larger values of $\alpha$, the FF and CR do not scale, although
the PSD and AF do; thus Eqs.~(\ref{fractalpsd}), and
(\ref{fractalaf}) imply each other for these FRSPPs.
Therefore, over the range of $\alpha$ for which the AF exhibits
scaling \cite{LOW96C}
\begin{equation}
\omega_0^\alpha T_1^\alpha =
\left\{ \begin{array}{ll}
\cos(\alpha \pi / 2) \Gamma(2 + \alpha) / (2 - 2^\alpha)
& \mbox{ for $0< \alpha < 1$}\\
\pi / \ln(4) & \mbox{ for $\alpha = 1$}\\
-\cos(\alpha \pi / 2) \Gamma(2 + \alpha) / (2^\alpha - 2)
& \mbox{ for $1 < \alpha < 3$.}
\end{array} \right.
\label{omeg}
\end{equation}
In principle, any of the these statistics could be solved for
$\alpha$ when it is known that $\alpha < 1$, although best results
obtain from the PG and the AF.

Since the GD of a fractal cannot exceed the Euclidean dimension,
which assumes a value of unity for one-dimensional point processes,
FSPPs cannot exist for $\alpha > 1$; only FRSPPs are possible at
these values of the fractal exponent.

\section{SYNTHESIS OF FRACTAL AND FRACTAL-RATE STOCHASTIC POINT PROCESSES}
\label{mathform}
In previous work we defined an FSPP and several FRSPPs and derived a
number of their statistics \cite{LOW95,LOW93A,LOW91}.
We include brief summaries of these here, as well as of the clustered
Poisson point-process model of Gr\"uneis and colleagues \cite{GRU86A}.
We also introduce two new fractal-rate processes (fractal lognormal
noise and fractal exponential noise) and define two new methods for
generating a point processes from a rate function:
integrate-and-fire and controlled variability integrate-and-fire.
These prove useful in isolating the effects of fractal components
from those of nonfractal noise introduced through the mechanics of
generating the point process.
For all of the processes considered, the scaling relation in
Eq.~(\ref{scale0}) holds for a number of statistical measures.

\subsection{Fractal Renewal Point Process (FRP)}
\label{frp_def}

The one-dimensional homogeneous Poisson point process (HPP) is
perhaps the simplest nonfractal stochastic point process \cite{HAI67}.
The HPP is characterized by a single constant quantity, its rate,
which is the number of events expected to occur in a unit interval.
A fundamental property of the HPP is that it is memoryless; given
this rate, knowledge of the entire history and future of a given
realization of a HPP yields no information about the behavior of the
process at the present.
The HPP belongs to the class of renewal point processes; times
between events are independent and identically distributed.
The HPP is the particular renewal point process for which this
distribution follows a decaying exponential form.

We now turn to a renewal process that is fractal: the standard
fractal renewal process (FRP) \cite{LOW93A,LOW92A,BER63,MAN65,LOW93B}.
In the standard FRP, the times between adjacent events are
independent random variables drawn from the same fractal
probability distribution.
In particular, the interevent-interval probability density function
$p(t)$ decays essentially as a power law function
\begin{equation}
\label{frp_sif}
p(t) =
\left\{ \begin{array}{ll}
k/t^{\alpha+1} & \mbox{ for $A < t < B$}\\
0 & \mbox{ otherwise,}
\end{array} \right.
\end{equation}
where $\alpha$ is the fractal exponent, $A$ and $B$ are cutoff
parameters, and $k$ is a normalization constant.
The FRP exhibits fractal behavior over time scales lying between $A$
and $B$.
This process is fully fractal for $0<\alpha<1$:
the power spectral density,
coincidence rate, Fano factor, Allan factor, and even the
interevent-time survivor function all exhibit scaling as in
Eq.~(\ref{scale0}) with the same power-law exponent $\alpha$ or
one simply related to it.

Further, for this process the capacity or box-counting dimension
$D_0$ assumes the value $\alpha$ \cite{LOW93A}; since the FRP is
ergodic, the generalized fractal dimension $D_q$ becomes
independent of the index $q$ \cite{THE90}, so
$D_q = D_0 = \alpha$ for all $q$, and all fractal
dimensions coincide.

A different (nonrenewal) point process results from the superposition
of a number $K$ of independent FRPs; however, for this combined
process as $K$ becomes large, and indeed for any FRSPP, the
interevent-time probability density function no longer scales, and
the generalized dimensions $D_q$ no longer equal $\alpha$, although
the PG and AF (and the FF and CR, for $0 < \alpha < 1$) retain their
scaling behavior.
As the number of such processes increases, and for certain ranges of
parameters, this superposition ultimately converges to the
fractal Gaussian-noise driven Poisson process \cite{LOW93A,LOW93B}
(see also Secs.~\ref{fgn} and \ref{fdspp}).

The standard FRP described above is a point process, consisting of a
set of points or marks on the time axis as shown in
Fig.~\ref{frp_diag}(a); however, it may be recast as a real-valued
process which alternates between two values, for example zero and
unity.
This alternating FRP would then start at a value of zero (for
example), and then switch to a value of unity at a time corresponding
to the first event in the standard FRP.
At the second such event in the standard FRP, the alternating FRP
would switch back to zero, and would proceed to switch back and forth
at every successive event of the standard FRP.
Thus the alternating FRP is a Bernoulli process, with times between
transitions given by the same interevent-interval probability density
as in the standard FRP, as portrayed in Fig.~\ref{frp_diag}(b).

\subsection{Integrate-and-Fire Fractal-Rate Stochastic Point Processes}
\label{fif}
Many point processes derive from a continuous-time function
$\lambda(t)$ which serves as the stochastically varying rate of the
point process.
These are known as fractal-rate stochastic processes (FRSPPs).
We confine our discussion to rate processes (and therefore FRSPPs)
which are ergodic.
Perhaps the simplest means for generating a point process from a rate
is the integrate-and-fire (IF) method \cite{TUR96}.
In this model, the rate function is integrated until it reaches a
fixed threshold $\theta$, whereupon a point event is generated and
the integrator is reset to zero.
Thus the occurrence time of the $(k+1)$st event is implicitly
obtained from the first occurrence of
\begin{equation}
\int_{t_{k}}^{t_{k+1}}\lambda(t)\,dt = \theta.
\label{ifdef}
\end{equation}
With such a direct conversion from the rate process to the point
process, any measure applied to these point processes will return
results closely related to the fractal structure of the underlying
rate process $\lambda(t)$.
In particular, over small frequencies, the theoretical
PSDs of the rate process and of the resulting point process
coincide.

We now turn to methods for generating several different kinds of
fractal rate functions.
More complex methods for point-process generation, as delineated in
Secs.~\ref{fjif} and~\ref{fdspp}, can also be applied to these same
rate functions.

\subsubsection{Fractal Gaussian Noise (FGN)}
\label{fgn}
Gaussian processes are ubiquitous in nature and are mathematically
tractable.
The values of these processes at any number of fixed times form a
jointly Gaussian random vector; this property, the mean, and the
spectrum completely define the process.
For use as a rate process, we choose a stationary process with a
mean equal to the expected rate of events, and a spectrum of the form
of Eq.~(\ref{fractalpsd}) with cutoffs.
FGN properly applies only for $\alpha < 1$; the range $1 < \alpha < 3$
is generated by fractal Brownian motion \cite{MAN83}.
However, in the interest of simplicity, we employ the term fractal
Gaussian noise for all values of $\alpha$.
A number of methods exist for generating FGN
\cite{FLA92A,MAN71,LUN86,STO94}: typically the result is a sampled
version of FGN with equal spacing between samples.
Interpolation between these samples is usually achieved by
selecting the most recently defined value.

With a rate process described by FGN serving as the input to an IF
process, the fractal-Gaussian-noise driven integrate-and-fire process
(FGNIF) results.
Generally, we require that the mean of the rate process be much
larger than its standard deviation, so that the times when the rate
is negative (during which no events may be generated) remain small.
The FGNIF has successfully been used to model the human heartbeat
\cite{TUR93,TUR96} (see Sec.~\ref{heartbeat}).

All of the rate processes considered in the following subsections
converge to FGN under appropriate circumstances, and thus the point
processes they yield through the IF construct converge to the FGNIF.

\subsubsection{Fractal Lognormal Noise (FLN)}
\label{fln}
A related process results from passing FGN through a memoryless
exponential transform.
Since the exponential of a Gaussian is lognormal, we call this
process fractal lognormal noise (FLN).
If $X(t)$ denotes an FGN process with mean ${\rm E}[X]$, variance
${\rm Var}[X]$, and autocovariance function $K_X(\tau)$, then
$\lambda(t) \equiv \exp[X(t)]$ is an FLN process with moments
${\rm E}[\lambda^n] = \exp\{n{\rm E}[X] + n^2{\rm Var}[X]/2\}$ and
autocovariance function
$K_\lambda(\tau) = {\rm E}^2[\lambda]\{\exp[K_X(\tau)] - 1\}$.

When this rate serves as the input to a doubly stochastic Poisson
point process (see Sec.~\ref{fdspp}), the result provides an
excellent model for vesicular exocytosis (see Sec.~\ref{exocytosis}).
By virtue of the exponential transform, the autocorrelation functions
of the Gaussian process $X(t)$ and the lognormal process $\lambda(t)$
differ; thus their spectra differ.
In particular, the PSD of the resulting point process does not follow
the exact form of Eq.~(\ref{fractalpsd}), although for small values
of ${\rm Var}[X]$, as encountered in modeling exocytosis (for which
${\rm Var}[X]\le 0.6$) \cite{LOW97A,LOW97B},
the experimental PG closely resembles this ideal form.
In the limit of very small ${\rm Var}[X]$, the exponential operation
approaches a linear transform, and the rate process becomes FGN.

\subsubsection{Fractal Exponential Noise (FEN)}
\label{fen}
Other nonlinear transforms may be applied to FGN, yielding other
fractal rate processes; we consider one which generates an
exponentially distributed process.
If $X_1(t)$ and $X_2(t)$ denote two independent and identically
distributed FGN processes with zero mean, variance ${\rm Var}[X]$,
and autocovariance function $K_X(\tau)$, then
$\lambda(t) \equiv X_1^2(t) + X_2^2(t)$ is a fractal exponential
noise (FEN) process with moments
${\rm E}[\lambda^n] = 2^n n! \left({\rm Var}[X]\right)^n$ and
autocovariance function $K_\lambda(\tau) = 4 K^2_X(\tau)$.
The rate $\lambda(t)$ turns out to be exponentially distributed.
If $K_X(\tau)$ scales as in Eq.~(\ref{fractalcr}) with an exponent
$\alpha_X$ in the range $1/2 < \alpha_X < 1$, then so will
$K_\lambda(\tau)$, but with an exponent
$\alpha_{\lambda} = 2 \alpha_X - 1$.

This process may prove useful in the study of certain kinds of
thermal light \cite{SAL78}.
Such light has an electric field which comprises two independent
components, with Gaussian amplitude distributions.
The light intensity is the sum of the squares of the two fields, and
therefore has an exponential amplitude distribution.

\subsubsection{Fractal Binomial Noise (FBN)}
\label{fbn}
A number of independent, identical alternating fractal renewal
processes (see Sec.~\ref{frp_def})
may be added together, yielding a binomial process with the
same fractal exponent as each individual alternating FRP
\cite{LOW93A}.
This binomial process can serve as a rate process for an
integrate-and-fire process; the fractal-binomial-noise driven
integrate-and-fire
(FBNIF) results.
It is schematized in Fig.~\ref{fbndp_dg}.
As the number of constituent processes increases, the FBN rate
converges to FGN (see Sec.~\ref{fgn}) with the same fractal exponent.

\subsubsection{Fractal Shot Noise (FSN)}
\label{fsn}
Though the HPP itself is not fractal, linearly
filtered versions of it, denoted shot noise, can exhibit fractal
characteristics.
In particular, if the impulse response function of the filter has a
decaying power-law (fractal) form, the result is fractal shot noise
(FSN) \cite{LOW89A,LOW89B,LOW90}.
If FSN serves as the rate for an IF process, the fractal-shot-noise
driven integrate-and-fire process (FSNIF) results;
Fig.~\ref{fsndp_dg} schematically illustrates the FSNDP as a
two-stage process.
The first stage produces a HPP with constant rate $\mu$.
Its output $M(t)$ becomes the input to a linear filter with a
power-law decaying impulse response function
\begin{equation}
h(t) =
\left\{ \begin{array}{ll}
k/t^{1-\alpha/2} & \mbox{ for $A < t < B$}\\
0 & \mbox{ otherwise,}
\end{array} \right.
\label{fsndp_ht}
\end{equation}
where $\alpha$ is a fractal exponent between zero and two,
$A$ and $B$ are cutoff
parameters, and $k$ is a normalization constant.
This filter produces fractal shot noise $I(t)$ at its output, which
then becomes the time-varying rate for the last stage, an
integrate-and-fire process.
The resulting point process $N(t)$ reflects the variations of the
fractal-shot-noise driving process.
Under suitable conditions, FSN converges to FGN as provided by the
central limit theorem \cite{LOW90}; the result is then the FGNIF.

\subsection{Controlled-Variability Integrate-and-Fire Fractal-Rate
\newline Stochastic Point Processes}
\label{fjif}
FRSPPs based on an integrate-and-fire substrate have only one source
of randomness, which is the rate process $\lambda(t)$.
Those based on a Poisson-process substrate (see Sec.~\ref{fdspp})
have a second source which depends explicitly on the rate process.
We now generalize the family of FRSPPs to include those which have a
second source of randomness that may be specified independently of
the first.
This new process differs from the simple IF process based on
$\lambda(t)$, by the imposition of a specific form of jitter on the
interevent times.
After generating the time of the $(n+1)$st event $t_{n+1}$ in
accordance with Eq.~(\ref{ifdef}), the $(n+1)$st interevent time
$t_{n+1} - t_n$ is multiplied by a Gaussian-distributed random
variable with unit mean and variance $\sigma^2$.
Thus $t_{n+1}$ is replaced by
\begin{equation}
t_{n+1} + \sigma (t_{n+1} - t_n) {\cal N}(0,1),
\end{equation}
where ${\cal N}(0,1)$ is a zero-mean, unity-variance Gaussian random
variable.
The result is the jittered-integrate-and-fire (JIF) family of
processes.
Employing the fractal-rate functions delineated in Sec.~\ref{fif}
yields the fractal-Gaussian-noise
driven jittered-integrate-and-fire process (FGNJIF) (Sec.~\ref{fgn}),
the fractal-lognormal-noise
driven jittered-integrate-and-fire process (FLNJIF) (Sec.~\ref{fln}),
the fractal-exponential-noise
driven jittered-integrate-and-fire process (FENJIF) (Sec.~\ref{fen}),
the fractal-binomial-noise
driven jittered-integrate-and-fire process (FBNJIF) (Sec.~\ref{fbn}),
and the fractal-shot-noise
driven jittered-integrate-and-fire process (FSNJIF) (Sec.~\ref{fsn}).

In these point processes, the standard deviation $\sigma$ of the
Gaussian jitter is a free parameter that controls the strength of
the second source of randomness.
Two limiting situations exist.
In the limit $\sigma \to 0$ the second source of randomness is absent
and the JIF processes reduce to the simple IF process considered in
Sec.~\ref{fif}.
The opposite limit, $\sigma \to \infty$ leads to a homogeneous Poisson
process; none of the fractal behavior in the rate process appears in
the point process, as if the rate were constant.
Between these two limits, as $\sigma$ increases from zero, the fractal
onset times of the resulting point processes increase, and the fractal
characteristics of the point processes are progressively lost,
first at higher
frequencies (shorter times) and subsequently at lower frequencies
(longer times).

Finally, we note that another generalization of the
integrate-and-fire formalism is possible \cite{TUR96}.
The threshold for firing $\theta$ need not remain constant, but can
itself be a random variable or stochastic process.
Randomness imparted to the threshold can then be used to represent
variability in other elements of a system, such as the amount of
neurotransmitter released per exocytic event or the duration of ion
channel openings.
The homogeneous Poisson process (HPP) with rate $\lambda_0$ would
then form a special case within this construct, generated when the
rate is fixed and constant at $\lambda(t) = \lambda_0$, and $\theta$
is an independent, exponentially distributed unit mean random
variable associated with each interevent interval.
If the rate process $\lambda(t)$ is also random, a doubly stochastic
Poisson point process results instead of the HPP.

\subsection{Fractal-Rate Doubly Stochastic Poisson Point Process\newline
(FRDSPP)}
\label{fdspp}
We have seen that an integrate-and-fire process (with
or without jitter) converts a fractal rate process into an FRSPP.
A Poisson process substrate may be employed instead of the IF
substrate, in which case the outcome is the family of doubly
stochastic Poisson point processes (DSPPs) \cite{COX55}.
DSPPs display two forms of randomness: one associated with the
stochastically varying rate, and the other associated with the
underlying Poisson nature of the process even if the rate were fixed
and constant.
With fractal rate processes, a fractal-rate DSPP (FRDSPP) results
\cite{LOW95}.
Again, simple relations exist between measures of the rate
$\lambda(t)$ and the point process $\partial N(t)$.
In particular, for the theoretical PSDs we have
\begin{equation}
S_N(\omega)= {\rm E}[\lambda] + S_{\lambda}(\omega),
\end{equation}
where $S_N(\omega)$ is the PSD of the point process and
$S_{\lambda}(\omega)$ is the PSD of the rate.

As with the fractal IF processes considered above, the amount of
randomness introduced by the second source in fractal DSPPs (the
Poisson transform) may also be adjusted.
However, in this case the amount introduced depends explicitly on the
rate $\lambda(t)$, rather than being independent of it as with the
JIF family of FRSPPs.
For example, for a particular rate function $\lambda(t)$ suppose that
its integral between zero and $t$ assumes the value $\Lambda$:
\begin{equation}
\int_0^t \lambda(u)\, du = \Lambda.
\end{equation}
Then given $\Lambda$, the number of events $N(t)$ generated in the
interval $(0,t]$ has a Poisson distribution, and in particular the
mean and variance of $N(t)$ both assume a value of $\Lambda$.
Therefore, if $\lambda(u)$ is multiplied by a constant value, then
for a fixed integration time $t$, the mean and variance of $N(t)$
will change by the same factor.
As this constant value increases, the coefficient of variation (the
standard deviation divided by the mean) of $N(t)$ decreases, and the
contribution of the Poisson transform to the total randomness of the
point process decreases, ultimately becoming negligible as the
multiplication constant increases without limit.
However, the DSPP formalism is less flexible than the JIF
formalism for introducing
randomness, since in the former case it explicitly depends on the
mean number of events, while in the latter case the variance may
be independently adjusted with the parameter $\sigma$.

Again, a variety of fractal point processes exist in the FRDSPP
class, depending on the rate process chosen.
Examples include
the fractal-Gaussian-noise driven Poisson point process
(FGNDP) (Sec.~\ref{fgn}, \cite{LOW91}),
the fractal-lognormal-noise driven Poisson point process
(FLNDP) (Sec.~\ref{fln}, \cite{LOW97A,LOW97B}),
the fractal-exponential-noise driven Poisson point process
(FENDP) (Sec.~\ref{fen}),
the fractal-binomial-noise driven Poisson point process
(FBNDP) (Sec.~\ref{fbn}, \cite{LOW93A,LOW95}),
and the fractal-shot-noise driven Poisson point process
(FSNDP) (Sec.~\ref{fsn}, \cite{LOW91}).
For the FGNDP a slight modification is necessary since a negative
rate has no meaning for a Poisson point process.
Therefore, rather than using a Gaussian process directly we instead
set the rate to zero for negative values of this process.
We require that the mean rate be much larger than its standard
deviation, so that the effect of this nonlinear limiting can
essentially be ignored.
Again, all forms of FRSPPs with positive finite mean rates converge
to the FGNDP under appropriate conditions \cite{LOW95}.

\subsection{Cluster Poisson Point Process}
\label{cluster}
Other important formulations for FRSPPs exist.
Gr\"uneis and colleagues defined a clustered Poisson point process in
which each member of a sequence of primary events from an HPP gives
rise to a train of secondary events (as in the FSNDP), but where the
events in the train form a segment of a renewal process (usually also
an HPP, but often with a different rate), with a fractal (power-law
distributed) duration \cite{GRU84,GRU86A}.
The resulting process indeed exhibits power-law scaling in the same
statistics as other FRSPPs \cite{GRU86B}.
The cluster Poisson point process is therefore a Bartlett-Lewis-type
process, whereas the FSNDP is a process of the Neyman-Scott type
\cite{COX80}.
This process has been used successfully to model mesencephalic reticular formation neural spike trains \cite{GRU93} (see Sec.~\ref{cns}).

\section{ESTIMATION OF THE ALLAN FACTOR AND POWER SPECTRAL DENSITY:
THEORY}
\label{sec:bias}
Given a segment of an FRSPP, we seek an estimate
$\widehat{\alpha}$ of the true fractal exponent $\alpha$ of the
entire process from which the segment was derived.
Several effects contribute to estimation error for finite-length
segments, regardless of the method used.
The fractal exponent provides a measure of the relative strengths of
fluctuations over various time scales; for an FRSPP with a relatively
large fractal exponent, for example, relatively more energy is
concentrated in longer time scale fluctuations than for an FRSPP with
a smaller fractal exponent.
Variance stems from the inherent randomness of the strengths of these
fluctuations.
A collection of finite realizations of an FRSPP with the same
parameters will exhibit fractal fluctuations of varying strengths,
leading to a distribution of estimated fractal exponents.

Bias appears to arise from cutoffs in the time (frequency) domain
which give rise to oscillations in the frequency (time) domain,
confounding the pure power-law behavior of the fractal PG (AF).
In addition, the physical limitations of the measurement process
itself impose practical limits on the range of time scales available.
Although algorithms exist for accurately compensating for these
cutoffs, they presuppose a detailed knowledge of the process {\it a
priori}, which is not in the spirit of estimating an unknown signal.
Consequently, in this paper we do not attempt to compensate for these
cutoffs in this manner.
In previous work \cite{LOW95} we determined the theoretical expected
Fano factor values for an FRSPP with finite duration, and the
corresponding expected bias for the corresponding fractal exponent.
Bias for the power spectral density was also considered, but not for
a finite data length.

The AF and the PG highlight the fractal behavior of FRSPPs
particularly well, and thus prove to be most useful for estimating
fractal exponents, as indicated in Sec.~\ref{analysis}.
We proceed to investigate the bias in these measures, and the effects
of this bias on fractal-exponent estimation.
The variance of PG-based estimators was examined previously
\cite{LOW95}; that for the AF appears not to be readily amenable to
analytical study.

\subsection{Effects of Finite Data Length on the AF Estimate}
\label{af:bias}
Unlike the Fano factor, estimates of the AF do not suffer from bias
due to finite data length; thus fractal exponent estimates based on
the AF do not either.
Given a particular data set, the estimated AF $\widehat{A}(T)$ at a
particular counting time $T$ is given by
\begin{equation}
\label{meana}
\widehat{A}(T) = \left. (N-1)^{-1}\sum_{k=0}^{N-2}(Z_{k+1}-Z_{k})^{2}
\right/ 2 N^{-1}\sum_{k=0}^{N-1}Z_{k},
\end{equation}
where $N$ is the number of samples,
$\{ Z_{k} \}$ represents the sequence of counts as shown in
Fig.~\ref{basics}(c), and the functional dependence of $Z_k$ upon $T$
is suppressed for notational clarity.

This estimate of the AF is simply the estimate of the Allan variance
divided by twice the estimate of the mean; however, computing the
expected value of this estimate is not straightforward.
We therefore employ the true mean rather than its estimate in
Eq.~(\ref{meana}).
This does not appreciably affect the result, since the error so
introduced remains a constant factor for all counting times, and so
cancels in power-law slope calculations where logarithms are used.
In any case, the estimate of the Allan variance exhibits far larger
variations than the estimate of the mean, so the fluctuations in the
AF estimate are dominated by the former.
Therefore, the expected value of the AF estimate becomes
\begin{eqnarray}
{\rm E}\left[ \widehat{A}(T) \right]
& = & {\rm E}\left\{
\left. \widehat{{\rm E}}\left[(Z_k - Z_{k+1})^2\right] \right/
2 \widehat{{\rm E}}[Z_k] \right\} \nonumber\\
& \approx & \left. {\rm E}\left[(Z_k - Z_{k+1})^2\right]
\right/ 2 {\rm E}[Z_k] \nonumber\\
& = & \left. {\rm E}\left[(Z_0 - Z_1)^2\right]
\right/ 2 {\rm E}[Z_0] \nonumber\\
& = & (\lambda T)^{-1} {\rm E}\left[Z_0^2 - Z_0 Z_1\right],
\label{afbiaseq}
\end{eqnarray}
which is independent of the number of samples, $N$, and
therefore of the duration of the recording $L$.
In the limit $L \to \infty$, the expected estimated AF for an FRSPP
with $0 < \alpha < 3$ follows the form of Eq.~(\ref{fractalaf})
precisely, by definition of an ergodic process.
Since Eq.~(\ref{afbiaseq}) does not depend on $L$, the quantity
${\rm E}\left[ \widehat{A}(T) \right]$ must assume a constant value
independent of $L$; thus the AF does not exhibit bias caused by
finite data duration.
Finally, since the expected value of the AF estimate is unbiased, the
expected value of the estimate of $\alpha$ itself is expected to have
negligible bias.
This is a simple and important result.

Estimates of the ordinary variance and Fano factor, in contrast, do
depend on the duration.
In this case we have
\begin{eqnarray}
\widehat{\rm Var}[Z]
& = & (N-1)^{-1}\sum_{k=0}^{N-1}(Z_{k}-\widehat{\rm E}[Z])^2
\nonumber\\
& = & (N-1)^{-1}\sum_{k=0}^{N-1}
\left(Z_{k}-N^{-1}\sum_{l=0}^{N-1}Z_{l}\right)^2 \nonumber\\
& = & (N-1)^{-1}\left[\sum_{k=0}^{N-1}Z_k^2
- 2N^{-1}\sum_{k=0}^{N-1}\sum_{l=0}^{N-1} Z_k Z_l
+ N^{-2}\sum_{k=0}^{N-1}\sum_{l=0}^{N-1}\sum_{m=0}^{N-1}
Z_l Z_m\right] \nonumber\\
& = & (N-1)^{-1}\sum_{k=0}^{N-1}Z_k^2
- N^{-1}(N-1)^{-1}\sum_{k=0}^{N-1}\sum_{l=0}^{N-1} Z_k Z_l \nonumber\\
& = & N^{-1}\sum_{k=0}^{N-1}Z_k^2
- N^{-1}(N-1)^{-1}\sum_k^{N-1}\sum_{l\neq k} Z_k Z_l.
\end{eqnarray}
These cross terms, which do depend on the number of samples, lead to
an estimated Fano factor with a confounding linear term
\begin{equation}
{\rm E}\left[\widehat{F}(T)\right] \approx 1 + (T/T_0)^\alpha
- T\left/\left(T_0^{\alpha} L^{1-\alpha}\right)\right.
\label{ffbias}
\end{equation}
for $0 < \alpha < 1$ \cite{LOW95}.
The last term on the right-hand-side of
Eq.~(\ref{ffbias}) leads to bias in estimating the
fractal exponent; for this and other reasons, we do not employ the
FF in fractal-exponent estimation.

\subsection{Effects of Finite Data Length on the PSD Estimate}
\label{fini}
Computation of the periodogram (estimated PSD) proves tractable only
when obtained from the series of counts $\{Z_k\}$, rather than
from the entire point process $N(t)$.
This, and more importantly the finite length of the data set,
introduce a bias in the estimated fractal exponent as shown below.

We begin by obtaining the discrete-time Fourier transform of the
series of counts $\{Z_k\}$, where $0 \le k < M$:
\begin{equation}
\tilde Z_n \equiv \sum_{k=0}^{M-1} Z_k e^{-j2\pi kn/M}.
\end{equation}
The periodogram then becomes
\begin{eqnarray}
\widehat{S}_Z(n)
& = & M^{-1} \left| \tilde Z_n \right|^2 \nonumber\\
& = & M^{-1} \sum_{k=0}^{M-1} \sum_{m=0}^{M-1}
Z_k Z_m e^{j2\pi (k-m)n/M},
\end{eqnarray}
with an expected value
\begin{eqnarray}
\E\left[\widehat{S}_Z(n)\right]
& = & M^{-1} \sum_{k=0}^{M-1} \sum_{m=0}^{M-1}
e^{j2\pi (k-m)n/M} \E\left[Z_k Z_m\right] \nonumber\\
& = & M^{-1} \sum_{k=0}^{M-1} \sum_{m=0}^{M-1} e^{j2\pi (k-m)n/M}
\E\left[\int_{s=0}^T\int_{t=0}^T\, dN(s+kT)dN(t+mT)\right] \nonumber\\
& = & M^{-1} \sum_{k=0}^{M-1} \sum_{m=0}^{M-1} e^{j2\pi (k-m)n/M}
\int_{s=0}^T\int_{t=0}^T G_N[s-t+(k-m)T]\, ds\, dt \nonumber\\
& = & M^{-1} \sum_{k=0}^{M-1} \sum_{m=0}^{M-1} e^{j2\pi (k-m)n/M}
\int_{u=-T}^T\int_{v=|u|}^{2T-|u|} G_N[u+(k-m)T]\,
\fracd{du\, dv}{2} \nonumber\\
& = & M^{-1} \sum_{k=0}^{M-1} \sum_{m=0}^{M-1} e^{j2\pi (k-m)n/M}
\int_{u=-T}^T(T-|u|)\, G_N[u+(k-m)T]\, du \nonumber\\
& = & M^{-1} \sum_{k=0}^{M-1} \sum_{m=0}^{M-1} e^{j2\pi (k-m)n/M}
\int_{u=-T}^T(T-|u|) \nonumber\\
&& \qquad \times \int_{\omega=-\infty}^\infty S_N(\omega)\,
e^{j\omega [u + (k-m)T]}\, \fracd{d\omega}{2 \pi}\, du \nonumber\\
& = & (2\pi M)^{-1} \int_{\omega=-\infty}^\infty S_N(\omega)
\left|\sum_{k=0}^{M-1} e^{jk(2\pi n/M + \omega T)}\right|^2
\int_{u=-T}^T(T-|u|)\, e^{j\omega u} \, du\, d\omega \nonumber\\
& = & (2\pi M)^{-1} \int_{\omega=-\infty}^\infty S_N(\omega)
\fracd{\sin^2(\pi n + M\omega T/2)}{\sin^2(\pi n/M + \omega T/2)}
\fracd{4\sin^2(\omega T/2)}{\omega^2}\, d\omega \nonumber\\
& = & \fracd{T}{\pi M} \int_{-\infty}^\infty S_N(2x/T)
\fracd{\sin^2(Mx)\sin^2(x)} {x^2\sin^2(x + \pi n/M)}\, dx.
\end{eqnarray}

For a general fractal stochastic point process, where the PSD follows
the form of Eq.~(\ref{fractalpsd}), we therefore have
\begin{eqnarray}
\E\left[\widehat{S}_Z(n)\right]
& = & \fracd{\lambda T}{\pi M} \int_{-\infty}^\infty
\left[1 + (\omega_0 T/2)^\alpha |x|^{-\alpha}\right]
\fracd{\sin^2(Mx)\sin^2(x)} {x^2\sin^2(x + \pi n/M)}\, dx.
\label{fractalpgex}
\end{eqnarray}
Focusing on the smaller values of $n$ useful in estimating fractal
exponents permits the use of two approximations in
Eq.~(\ref{fractalpgex}).
For the values of $\alpha$ used in this paper ($0 < \alpha < 2$), the
integrand of Eq.~(\ref{fractalpgex}) will only be significant near
$x = -\pi n/M$, yielding
\begin{eqnarray}
\E\left[\widehat{S}_Z(n)\right]
& \approx & \fracd{\lambda T}{\pi M} \int_{-\infty}^\infty
\left[1 + (\omega_0 T/2)^\alpha |x|^{-\alpha}\right]
\fracd{M\pi\delta(x + \pi n/M)\sin^2(x)} {x^2}\, dx \nonumber\\
& = & \lambda T \left[1 + (2\pi n / \omega_0 M T)^{-\alpha}\right]
\fracd{\sin^2(\pi n/M)} {(\pi n/M)^2} \nonumber\\
& \approx & \lambda T
\left[1 + (2\pi n / \omega_0 M T)^{-\alpha}\right],
\end{eqnarray}
which is of the same form as Eq.~(\ref{fractalpsd}).
Improvement of this estimation procedure appears to require
numerical integration of Eq.~(\ref{fractalpgex}), which proves
nontrivial since the integrand exhibits oscillations with a small
period $\pi/M$.
Fortuitously, for the parameter values employed in this paper the
integrand appears peaked near $x = -\pi n/M$, so that not too many
($\approx 200$) oscillations need be included in the calculations.
Indeed, numerical results employing this method, and with $\omega_0
\to 0$ for which Eq.~(\ref{fractalpgex}) is known to assume the
simple value $\lambda T$, agree within $0.1\%$.
These results form an extension of earlier approaches
\cite{LOW95,LOW93D} which ignored the effects of imposing periodic
boundary conditions on the Fourier transform, and of binning the
events.
Numerical integration of Eq.~(\ref{fractalpgex}) followed by a
least-squares fit on a doubly logarithmic plot leads to results for
the expected bias of the PSD-based estimate of the fractal exponent,
shown in Table~\ref{tabgau02}.

Other methods exist for compensating for finite data length in power
spectral estimation, such as maximizing the entropy subject to the
correlation function over the available range of times
(see, e.g., \cite{SKIL88}).

\section{ESTIMATION OF THE ALLAN FACTOR AND POWER SPECTRAL DENSITY:
SIMULATION}
\label{methodsand}
Having considered various possible theoretical sources of error in
simulating FRSPPs with a desired fractal exponent, we now proceed to
investigate their effects upon fractal exponent estimates based on
the power spectral density and Allan factor measures.
To this end we employ simulations of three of the FRSPPs outlined
above: FGN driving a deterministic integrate-and-fire process
(FGNIF), FGN driving an IF process followed by Gaussian jitter of the
interevent intervals (FGNJIF), and FGN driving a Poisson process
(FGNDP).
We choose these three processes from the collection presented in
Sec.~\ref{mathform} for a number of reasons.
First, the FGNIF, FGNJIF, and FGNDP derive from a common
continuous-time rate process, fractal Gaussian noise (FGN), and thus
differences among these point processes must lie in the point-process
generation mechanism rather than in the fractal rate processes.
Second, FGN admits fractal exponent of any value, while the fractal
exponents cannot exceed two for fractal binomial noise and fractal
shot noise.
Third, of all the FRSPPs examined in Sec.~\ref{mathform}, those
based on the FGN appear to suffer the least from the effects of
cutoffs, so that expected values of the PG and the AF most closely
follow the pure power-law forms of Eqs.~(\ref{fractalpsd})
and~(\ref{fractalaf}) respectively.

To generate the rate function with $1/f^{\alpha}$ spectral
properties, we computed discrete-time approximations to fractal
Gaussian noise (FGN) by forming a conjugate-symmetric sequence
$\tilde{X}[k]$ of $M$ samples which satisfies
\begin{equation}
\left|\tilde{X}[k]\right|=\left\{
\begin{array}{ll}
M \rho & \mbox{for $k=0$}\\
c k^{-\alpha/2} & \mbox{for $1\leq k \leq M/2$},
\end{array}
\right.
\end{equation}
where $\rho = {\rm E}[\lambda(t)]$ is the desired average value of
the FGN process, and $c$ is a constant determining the relative
strength of the FGN.
The phases of $\tilde{X}[k]$ were independent and uniformly
distributed in the interval $[0,2\pi)$ for $1\leq k < M/2$, while both
$\tilde{X}[0]$ and $\tilde{X}[M/2]$ were strictly real.
A discrete-time sequence $\tilde{x}[n]$ was obtained by taking the
inverse discrete Fourier transform of $\tilde{X}[k]$.
The resulting FGN process is periodic with period $M$, so to reduce
the effects of this periodicity we employ only the first half of the
sequence.
Specifically, we used an original sequence $\tilde{x}[n]$ of length
$M = 2^{16}=65\,536$ points and kept the first $2^{15}= 32\,768$
samples as $x[n]$.
Further, since rate functions must not attain negative values, we
chose the mean value of the FGN, $\rho$, sufficiently large compared
to the relative strength of the FGN $c$ to ensure that $x[n]$
remained positive for all $2^{15}$ different values of $x[n]$ in all
simulations.
In particular for $\alpha < 1$ this was adjusted by choosing the FF
intercept time $T_0$ to be 10 times the average interevent time,
while for $\alpha > 1$ the PSD fractal onset frequency $\omega_0$ was
chosen to be $0.0005$ times the average rate $\rho$.

The resultant sequence $x[n]$ was taken to represent equally
spaced samples of a con\-tin\-u\-ous-time process;
to simplify subsequent analysis, we chose the duration of each sample
to be one sec, without loss of generality.
Varying the mean event production rate $\rho$ while keeping all other
parameters fixed (or in fixed relation to $\rho$, as with the FF
intercept time and PSD corner frequency) resulted in point
processes which contained varying expected numbers of events.
In particular, we employed values of $\rho = 5, 10, 20, 40$ and
$80$ events per sample of $x[n]$, leading to a range of roughly
$164\,000$ to $2\,620\,000$ events in the resulting point process.

The following procedures were employed to compute the estimates
described in Sec.~\ref{analysis}.
For the PG, the absolute event times were quantized into
$2^{16}$ bins, which therefore formed a rate estimate of the process
with counting windows having a duration of $0.5$ sec.
Then a discrete Fourier transform was performed, followed by
replacing the Fourier components with their square magnitudes.
Finally, a least-squares fit was performed on the logarithm of these
quantities vs.\ the logarithm of the frequencies $f$ for various
selected ranges of frequencies.
These slopes form PG-based estimates of the fractal exponent.

Estimates of the AF were calculated over a range of counting times
$T$ in a geometric sequence with ten counting times per decade.
The slopes of the AF estimate were again obtained from linear
least-squares fits to the logarithm of the AF values vs.\ the
logarithm of the size of the counting time $T$; these slopes became
the AF-based estimates of the fractal exponent.
As indicated in Sec.~\ref{af:bias}, finite data lengths do not
impart extraneous bias to the AF-based estimates.

For each type of FRSPP and value of $\rho$ employed, 100 simulations
were performed for each of three different design values of the
fractal exponent: $\alpha=0.2$, 0.8, and 1.5; AF and PG curves
were computed for each simulation.
Estimates of the fractal exponent were obtained by two methods for
each of these two measures.
In the first, all 100 AF or PG curves were averaged together, after
which the doubly logarithmic slope of this average curve was
computed, yielding a single estimate of the fractal exponent.
The second method consisted of first obtaining the doubly logarithmic
slopes of
the individual AF or PG curves, and then averaging these slopes.
This second method yields the mean fractal exponent as well as the
standard deviation (and other possible statistics of the
fractal exponent).

As shown in Sec.~\ref{fif}, the FGN-driven integrate-and-fire
process (FGNIF) is obtained by identifying the FGN sequence $x[n]$ as
samples of a rate function $\lambda(t)$, taken to be constant over
each sample of $x[n]$ (1 sec).
This rate function is integrated until the result attains a value of
unity, whereupon an output point is generated and the integrator
is reset to zero.
The generation of the continuous rate function from the
piecewise-constant FGN does not significantly change the observed
fractal structure of the resulting point process.
For frequencies less than the inverse of the sampling time $T$ (1 Hz
in this example), the PSDs of the piecewise-constant and exact
versions of FGN differ by a factor of
$\sinc^2(\omega T/2)$; fractal behavior depends on low
frequencies ($\omega \ll 1/T$) where the above factor
is essentially unity.
Since the FGNIF process has just one source of randomness, namely the
underlying FGN, the estimators for the fractal exponent should
therefore display only the (fractal) behavior of the FGN itself,
along with finite-size effects.
Bias or variance due to a second source of randomness, such as the
Poisson transform in the FGNDP, will not be present.
Thus for our purposes the FGNIF process serves as a benchmark for the
accuracy of fractal exponent estimators, with a bias and variance
depending only on finite data size effects and not on the rate
$\rho$.

The results for the FGNIF process with a rate $\rho = 40$ and three
different fractal exponents ($\alpha$=0.2, 0.8 and 1.5) are
summarized for the AF in Table~\ref{tabgau01} and for the PG in
Table~\ref{tabgau02}.
\begin{table}[tbph]
\begin{center}
\begin{tabular}{|c|c|l|l|l|}
\hline
Range (sec.) & Method  & \multicolumn{1}{c|}{$\alpha=0.2$} &
\multicolumn{1}{c|}{$\alpha=0.8$} & \multicolumn{1}{c|}{$\alpha=1.5$}\\
\hline
$62.5-625$ &Fit of Avg.\ & $0.199$         & $0.799$         & $1.499$\\
 &          Avg.\ of Fits & $0.194\pm0.074$ & $0.795\pm0.072$ & $1.495\pm0.067$\\
\hline
$125-1250$ &Fit of Avg.\ & $0.211$         & $0.807$         & $1.499$\\
 & Avg.\ of Fits & $0.199\pm0.119$ & $0.807\pm0.114$ & $1.490\pm0.105$\\
\hline
$250-2500$ & Fit of Avg.\ &$0.219$         & $0.804$         & $1.490$\\
 &   Avg.\ of Fits & $0.184\pm0.165$ & $0.804\pm0.153$ & $1.472\pm0.138$\\
\hline
$25-2500$ &Fit of Avg.\ & $0.204$         & $0.801$         & $1.495$\\
 & Avg.\ of Fits & $0.192\pm0.056$ & $0.792\pm0.057$ & $1.487\pm0.059$\\
\hline
\end{tabular}
\end{center}
\caption{
AF-based fractal exponent estimates $\widehat{\alpha}$ for simulated
FGNIF processes with a rate $\rho=40$, for different time ranges.
The governing rate processes have theoretical fractal exponents of
$\alpha=0.2$, $0.8$, and $1.5$.
The estimated fractal exponents were obtained from straight-line fits
on doubly logarithmic coordinates to an average curve of 100
independent simulations (Fit of Avg.) and from averages of the slopes
of the individual curves (Avg.\ of Fits).
The deviations are the standard deviations obtained from the second
averaging procedure.}
\label{tabgau01}
\end{table}
\begin{table}[tbph]
\begin{center}
\begin{tabular}{|c|c|l|l|l|}
\hline
Range (Hz) & Method &\multicolumn{1}{c|}{$\alpha=0.2$} &
\multicolumn{1}{c|}{$\alpha=0.8$} & \multicolumn{1}{c|}{$\alpha=1.5$}\\
\hline
$0.00025 - 0.0025$
& Fit of Avg.\ & $0.200$         & $0.835$         & $1.628$\\
 &Avg.\ of Fits &
 $0.201\pm0.140$ & $0.836\pm0.137$ & $1.623\pm0.118$\\
& Theory &
 $0.200\pm0.116$ & $0.807\pm0.116$ & $1.545\pm0.116$\\
\hline
$0.0005 - 0.005$
& Fit of Avg.\ & $0.196$         & $0.822$         & $1.614$\\
& Avg.\ of Fits &
 $0.197\pm0.094$ & $0.823\pm0.093$ & $1.610\pm0.088$\\
& Theory &
 $0.200\pm0.082$ & $0.804\pm0.082$ & $1.535\pm0.082$\\
\hline
$0.001 - 0.01$
& Fit of Avg.\ & $0.202$         & $0.821$         & $1.605$\\
& Avg.\ of Fits &
 $0.202\pm0.078$ & $0.821\pm0.077$ & $1.602\pm0.075$\\
&Theory &
 $0.200\pm0.058$ & $0.803\pm0.058$ & $1.526\pm0.058$\\
\hline
$0.002 - 0.02$
& Fit of Avg.\ & $0.206$         & $0.828$         & $1.594$\\
& Avg.\ of Fits &
 $0.202\pm0.059$ & $0.824\pm0.057$ & $1.589\pm0.057$\\
&Theory &
 $0.201\pm0.041$ & $0.802\pm0.041$ & $1.520\pm0.041$\\
\hline
 $0.0002 - 0.02$
& Fit of Avg.\ & $0.206$         & $0.831$         & $1.609$\\
& Avg.\ of Fits &
 $0.204\pm0.030$ & $0.830\pm0.031$ & $1.604\pm0.043$\\
&Theory &
 $0.201\pm0.039$ & $0.803\pm0.039$ & $1.527\pm0.039$\\
\hline
\end{tabular}
\end{center}

\caption{
PG-based fractal exponent estimates
$\widehat{\alpha}$ for simulated FGNIF processes with a rate
$\rho=40$, for different frequency ranges.
The governing rate processes have theoretical fractal exponents of
$\alpha=0.2$, $0.8$, and $1.5$.
The estimated fractal exponents were obtained from straight-line fits
in doubly logarithmic coordinates to an average curve of 100
independent simulations (Fit of Avg.) and from averages of the slopes
of the individual curves (Avg.\ of Fits).
The deviations are the standard deviations obtained from the second
averaging procedure.
Also included are the theoretical bias values obtained from
Eq.~(32) and standard deviation values from [2, Eq.~(17)].
These standard deviations are in substantial agreement with the data.
Predictions of the bias underestimate the magnitude by factors of
3--10, although they correctly predict the sign.
}
\label{tabgau02}
\end{table}
These Tables present estimated values of the fractal exponent
$\widehat{\alpha}$ obtained from averages over 100 independent
simulations, for a variety of time and frequency ranges.
The exponents were obtained by the two methods delineated above, and
are labeled ``Fit of Avg.'' and ``Avg.\ of Fits'', respectively;
theoretical results for the PSD are labeled ``Theory''.
For an experimental AF or PG following the form of
Eq.~(\ref{fractalaf}) or~(\ref{fractalpsd}), respectively, the range
of times or frequencies over which the fractal exponent is estimated
will not significantly affect the returned value, as long as the
range lies in the scaling region $\omega_0^{-1} \ll T \ll L$ (or
$1/L \ll \omega \ll \omega_0$), respectively.

To demonstrate this, we estimated the fractal exponents over four
time ranges (for the AF) and five frequency ranges (for the PG), with
each range spanning a factor of ten or one hundred.
As shown in Tables~\ref{tabgau01} and~\ref{tabgau02}, the results for
the averaged AF and PG curves do not depend significantly on the
choice of fit ranges, and in general agree well with the design
values of the fractal exponents $\alpha$.
The deviations of the PG-based fractal exponent estimator
from these design values are accounted for in part by the theoretical
bias values obtained from the general result provided in
Eq.~(\ref{fractalpgex}), as shown in
Table~\ref{tabgau02}; these theoretical predictions yield the correct
sign for the bias, but underestimate the magnitude by factors of
3--10.
Employing predictions directly based on the FGNIF process (see
Appendix), rather than on a general FRSPP, yield results in
substantial agreement with those of Eq.~(\ref{fractalpgex}).
The AF-based fractal exponent estimators, in contrast, do not suffer
from bias due to finite data size, and thus agree with the design
values directly.
In particular, the AF yields better performance than the PG estimate,
especially for the larger design fractal exponents values of
$\alpha=0.8$ and $1.5$.

The fractal exponents computed by the second method are also listed
in Tables~\ref{tabgau01} and~\ref{tabgau02}, indicated by ``Avg.\ of
Fit'', together with their standard deviations.
(Note that for the first method there is no direct procedure to
calculate the standard deviation since the results are extracted from
a single curve).
The two methods for estimating the fractal exponent agree remarkably
well.
For an ordinary Gaussian random variable, averaging over 100 samples
would yield estimated values of the mean which exceeded the true mean
by $2/10$ of a standard deviation 5\% of the time, with the rest
clustered more tightly about the true mean.
Of the 27 experiments in which both methods for estimating the
fractal exponent were performed, for only one (AF, $\alpha=0.2$, and
$250<T<2500$) does the sample mean exceed this limit, suggesting that
this simple Gaussian model is valid.

Useful estimators must have small variance as well as minimal bias;
we therefore now focus on the standard deviations in
Tables~\ref{tabgau01} and~\ref{tabgau02}.
For both the AF- and PG-based estimators, and for ranges spanning one
decade, the standard deviation decreases slightly as the design
values of the fractal exponent increase; for two-decade ranges, this
trend is reversed.
Standard deviations for the two-decade ranges are significantly less
than those for a single decade, as expected.
For ranges spanning one decade, the standard deviation increase with
decreasing frequency (PG) or increasing counting time (AF); at these
time and frequency scales, the PG has the worst proportional
frequency resolution, and the AF suffers from relatively few counting
windows over which to obtain accurate statistics.
Finally, for similar time and frequency ranges the AF exhibits larger
standard deviation than the PG.
This is also likely due to the limited number of counting windows
available, which proves more significant than the reduced resolution
of the PG.
Indeed, predictions of the estimated standard deviation for the PG
based on earlier work \cite{LOW95} agree well with the data; these
predictions in turn depend directly on the number of PG values
available.

A figure of merit for fractal exponent estimators which combines the
bias and variance is the root-mean-square error, equivalent to the
square root of the sum of the variance and the squared bias.
Employing the results for the individual AF and PG entries provided in
Tables~\ref{tabgau01} and~\ref{tabgau02} yields the results shown in
Table~\ref{rmse}.
\begin{table}[tbph]
\begin{center}
\begin{tabular}{|c|c|c|c|c|}
\hline
Statistic &  Range & $\alpha=0.2$ & $\alpha=0.8$ & $\alpha=1.5$\\
\hline
AF & $62.5-625$      sec & $0.074$ & $0.072$ & $0.067$\\
   & $125-1250$      sec & $0.119$ & $0.114$ & $0.105$\\
   & $250-2500$      sec & $0.166$ & $0.153$ & $0.141$\\
   & $25-2500$       sec & $0.057$ & $0.058$ & $0.060$\\
\hline
PG & $0.00025-0.0025$ Hz & $0.140$ & $0.142$ & $0.170$\\
   & $0.0005-0.005$   Hz & $0.094$ & $0.096$ & $0.141$\\
   & $0.001-0.01$     Hz & $0.078$ & $0.080$ & $0.127$\\
   & $0.002-0.02$     Hz & $0.059$ & $0.062$ & $0.106$\\
   & $0.0002-0.02$    Hz & $0.030$ & $0.043$ & $0.113$\\
\hline
\end{tabular}
\end{center}
\caption{
Root-mean-square (rms) error of the estimated fractal exponent
$\widehat{\alpha}$ for both AF- and PG-based estimators, for the same
ranges and parameters used in Tables~1 and~2.
}
\label{rmse}
\end{table}
For both the AF- and PG-based estimates, the best performance is
associated with times and frequencies well away from the duration of
the data set: smallest lower counting time limits for the AF, and
largest upper frequency limits for the PG, respectively.
Comparing conjugate time and frequency ranges shows that the
fractal-exponent estimator based on the AF yields significantly
superior performance for the largest design value of $\alpha$, while
the PG proves somewhat better at the smallest value.
Exact relationships between cutoff frequencies and cutoff times
depend on the value of $\alpha$ itself \cite{LOW95}
[see also Eq.~(\ref{omeg})], so direct comparisons between ranges for
the AF- and PG-based estimates vary among the columns of
Table~\ref{rmse}.

The results obtained are governed by the finite duration of the data
and by the nature of the estimation process, and are not an artifact
of the particular value of the rate, $\rho$, that was chosen.
Reducing the rate from $\rho=40$ to $\rho=10$, while keeping all
other parameters fixed, generates results which differ from those
presented in Tables~\ref{tabgau01} and~\ref{tabgau02} by a maximum of
$\pm0.001$.
Thus in the FGNIF, the integrate-and-fire algorithm appears to
translate the fractal quality of the continuous FGN into a
point process nearly perfectly.
All differences between the estimated fractal exponents and the
design values used in the generation of the FGNIF simulations must
therefore be due to finite-size effects: bias for the PG-based
estimator, as discussed in Section~\ref{fini}, and variance for both PG-
and AF-based estimators.

We now proceed to examine other point processes based on FGN, such as
the FGNIF with jitter of the interevent intervals (FGNJIF) and the
FGNDP.
These processes, which include a second source of randomness
that is inherent in the mechanism of
point-process generation, exhibit increased fractal-exponent
estimation error.
We present averaged AF and PG plots for the FGNJIF with a mean rate
$\rho=10$ and jitter parameters $\sigma=0$, $0.5$, and $1$ in
Fig.~\ref{gausse}.
These plots attain asymptotic values at larger times (for the AF) and
lower frequencies (PG) than those for the FGNIF do, especially for
larger values of the jitter parameter $\sigma$.
This reduced range of scaling behavior serves to reduce the measured
slopes of these curves and thus the estimated values of the fractal
exponent for both the AF- and PG-based estimators, for all design
values of $\alpha$ employed.
For $\sigma=0$, corresponding to zero jitter (the FGNIF), the slopes
attain the values shown in Tables~\ref{tabgau01} and~\ref{tabgau02}.
Estimated fractal exponent values for the FGNJIF are presented in
Table~\ref{ifallsig}, which highlights the increasing negative bias
for increased $\sigma$.
\begin{table}[tbph]
\begin{center}
\begin{tabular}{|c|c|c|c|c|c|}
\hline
Statistic & Range             & Jitter $\sigma$ &
$\alpha=0.2$ & $\alpha=0.8$ & $\alpha=1.5$\\
\hline
AF& $250 - 2500$ sec & $0$   & $0.213$ & $0.804$ & $1.490$\\
  &		      & $0.5$ & $0.193$ & $0.796$ & $1.469$\\
  &		      & $1$   & $0.153$ & $0.779$ & $1.416$\\
\hline
PG&  $0.002 - 0.02$ Hz & $0$   & $0.206$ & $0.828$ & $1.594$\\
  &		      & $0.5$ & $0.189$ & $0.796$ & $1.104$\\
  &		      & $1$   & $0.143$ & $0.718$ & $0.649$\\
\hline
\end{tabular}
\end{center}
\caption{
Fractal exponents estimated from the PG and AF of the FGNJIF process,
with $\sigma$ chosen to be 0, 0.5, and 1.
The governing rate processes have design fractal exponents of
$\alpha=0.2$, $0.8$, and $1.5$, and mean values of $\rho = 10$ in all
cases.
One hundred individual curves were averaged, after which a single
exponent was computed.
As $\sigma$ increases the estimators exhibit an increasingly negative
bias.
}
\label{ifallsig}
\end{table}

The AF curves in Fig.~\ref{gausse} exhibit two additional
jitter-dependent features worthy of mention.
First, as $\sigma$ increases the regularity of the signal decreases,
leading to increased relative variance and a consequent decrease in
the amplitude of the dip near $T=1$ sec.
Second, while the AF must approach an asymptotic value of unity for
counting times $T$ less than half the minimum interevent interval
[see Eqs.~(\ref{afdef}) and~(\ref{afcalc})], this does not appear to
occur in Fig.~\ref{gausse} for $\sigma = 1$.
This apparent conundrum is resolved by recognizing that large values
of the jitter parameter $\sigma$ lead to an increased probability
that two adjacent event times will be perturbed to new values very
close to each other.
Thus as $\sigma$ increases, the relative frequency of quite small
interevent intervals also increases, and the AF must be extended to
ever smaller counting times to approach its limit of unity.
We chose not to plot the AF curves at these smaller counting times
$T$, since our focus lies in fractal behavior which manifests itself
over longer time scales.

We now turn to the FGNDP, which has a Poisson process substrate.
As the average rate $\rho$ decreases, the relative contribution of
the randomness introduced by the Poisson generator increases
as $1/ \sqrt{\rho} $, and the
PG- and AF-based fractal-exponents estimators exhibit increasingly
larger bias.
We present results for simulated FGNDP processes in
Fig.~\ref{poissons}; Table~\ref{prateall} provides estimated fractal
exponent values.
The average differences between the estimated values of the
fractal-exponents $\widehat{\alpha}$ and the design values $\alpha$
are shown graphically in Fig.~\ref{fehler} as a function of the mean
event production rate $\rho$.
\begin{table}[tbph]
\begin{center}
\begin{tabular}{|c|c|c|c|c|c|}
\hline
Statistic & Range & Rate $\rho$
& $\alpha=0.2$ & $\alpha=0.8$ & $\alpha=1.5$\\
\hline
AF& $250 - 2500$ sec &  $5$ & $0.157$ & $0.774$ & $1.462$\\
  &                  & $20$ & $0.153$ & $0.794$ & $1.474$\\
  &	     & $40$ & $0.197$ & $0.798$ & $1.483$\\
  &	     & $80$ & $0.185$ & $0.802$ & $1.489$\\
\hline
PG& $0.002 - 0.02$ Hz &  $5$ & $0.130$ & $0.652$ & $0.997$\\
  & 		      & $20$ & $0.132$ & $0.764$ & $1.092$\\
  & 		      & $40$ & $0.149$ & $0.786$ & $1.351$\\
  & 		      & $80$ & $0.157$ & $0.804$ & $1.496$\\
\hline
\end{tabular}
\end{center}
\caption{
Fractal exponents estimated from the PG and AF of the FGNDP process,
with a rate $\rho$ chosen to be 5, 20, 40, and 80.
The governing rate processes have design fractal exponents of
$\alpha=0.2$, $0.8$, and $1.5$.
One hundred individual curves were averaged, after which a single
exponent was computed.
As $\rho$ decreases the estimators exhibit an increasingly negative
bias.
}
\label{prateall}
\end{table}

For both the FGNJIF and the FGNDP processes, estimating fractal
exponents over larger ranges of time or frequency leads to increased
bias, particularly for large values of $\sigma$ and small values of
$\rho$.
Since the results depart from those obtained with the FGNIF, the
additional randomness as $\rho$ decreases, or as $\sigma$ increases,
leads to a dilution of the power-law behavior of the AF and PG,
changing its onset to at larger times and smaller frequencies.
In the limits $\rho \to 0$ and $\sigma \to \infty$, the result is
essentially white noise and the estimated fractal exponent will be
zero.

\section{DISCUSSION}

Best results for the FGNIF were obtained with the AF-based
fractal-exponent estimator using counting times in the range
$25 \le T \le 2500$; this yielded root-mean-square error
results which were accurate to within $0.06$ for all design values of
$\alpha$ (Tables \ref{tabgau01} and \ref{rmse}).
For both the AF- and PG-based estimators, performance degraded at
longer times and smaller frequencies
(Tables \ref{tabgau01}, \ref{tabgau02}, and \ref{rmse}).
Finite-duration effects in the PG and relatively few counting windows
in the AF produce significant error over these scales.
The rms error remained near $0.1$ for both estimators, over all design
values of the fractal exponent, and over all time and frequency
ranges examined.

The FGNJIF and FGNDP processes impose a second source of randomness
on the fractal behavior of the
FGN kernel.
This additional randomness is essentially uncorrelated, and
thus may be modeled as the linear superposition of white noise on the
original fractal signal; this affects the shapes of both the AF and
the PG.
As this white-noise level increases, it dilutes the fractal
structure present in the FGN process, beginning at the shortest
time scales and moving progressively towards longer ones.
This effect reduces the range of times and frequencies over which
fractal behavior appears, resulting in a decreased estimated fractal
exponent.
This proved most significant for larger design values of $\alpha$,
and for the estimator based on the PG.
Indeed, magnitudes of the bias exceeded $0.8$
in one case.
The AF-based estimator fared significantly better, although in this
case it employed counting times which were five times the inverse of
the frequencies used for the PG.
These results compare favorably to earlier studies using the same
number of events \cite{LOW95}, for which inaccuracies of $0.1$ or
larger were obtained for a variety of fractal stochastic point
processes, all of which included a second source of randomness.

This study considered data sets with an average of $10^6$ points;
experimental data may contain far fewer points, and thus yield poorer
fractal-exponent estimation statistics.
In particular, finite-size effects, which lead to increased bias for
the PG-based estimator and increased variance for the AF-based
estimator will reduce the ranges of times and frequencies over which
reliable estimation can be achieved.
Similarly, data sets with larger numbers of points (computer network
traffic, for example) would yield superior performance.
In addition, if several independent data sets are available, known to
be from identical experiments, then the AF plots or the resulting
estimates could be averaged, reducing the effective variance.
In this case, the reduced bias in the AF-based estimator would render
it far superior to that based on the PG.

Finally we mention that there exist several methods for reducing the
bias of the fractal exponent estimates if one has some {\it a
priori} knowledge of the process of interest.
Such an approach however does not follow the philosophy of estimating
a completely unknown signal and we therefore did not attempt to
compensate for the bias by the use of such methods.

\section{CONCLUSION}
We have investigated the properties of fractal and fractal-rate
stochastic point processes (FSPPs and FRSPPs), focusing on the
estimation of measures that reveal their fractal exponents.
The fractal-Gaussian-noise driven integrate-and-fire process (FGNIF)
is unique as a point process in that it exhibits no short-term
randomness; we have developed a generalization which includes jitter,
the FGNJIF, for which the short-term randomness is fully adjustable.
In addition to the randomness contributed by the fractal nature of
the rate, all other FRSPPs appear to exhibit additional
randomness associated with point generation,
which confounds the accuracy of fractal-exponent estimation.
The FGNJIF proved crucial in elucidating the role that such
randomness plays in the estimation of fractal exponents in other
FRSPPs.
We presented analytical results for the expected biases of the PG- and
AF-based fractional exponent estimators due to finite data length.
We followed this theoretical work with a series of simulations of
FRSPPs for three representative fractal exponents: $\alpha=0.2$,
$0.8$, and $1.5$.
Using these simulations together with the analytical predictions, we
delineated the sources of error involved in estimating fractal
exponents.
In particular, using the FGNJIF we were able to separate those
factors intrinsic to estimation over finite FRSPP data sets from those
due to the particular form of FRSPP involved.
We conclude that the AF-based estimate proves more reliable in
estimating the fractal exponent than the PG-based method, yielding
rms errors of $0.06$ for data segments of FRSPPs with $10^6$ points over
all values of $\alpha$ examined.
Finally, we note that wavelet generalizations of the AF appear
to yield comparable results \cite{TEI96A,HEN96}, suggesting that the AF
estimate may be optimal in some applications.

\section{ACKNOWLEDGMENTS}
We are grateful to M.~Shlesinger for valuable suggestions.
This work was supported by
the Austrian Science Foundation
under Grant No.\ S-70001-MAT (ST, MF, HGF),
by the Office of Naval Research
under Grant No.\ N00014-92-J-1251 (SBL, MCT),
by the Whitaker Foundation (SBL),
and by the Joint Services Electronics Program
through the Columbia Radiation Laboratory (MCT).

\section*{APPENDIX: PSD FOR THE FGNIF PROCESS}

Equation~(\ref{fractalpgex}) yields predictions for the PG-based
fractal exponent estimation bias for a general fractal-rate
stochastic point process (FRSPP).
More accurate results may be obtained for particular FRSPPs, taking
into account the structure of these individual processes.
We consider the case of the FGNIF process, with the further
specialization that the original FGN array is obtained by direct
Fourier synthesis with half of the array discarded, and that the
analyzing periodogram subsequently doubles the number of points,
yielding the same number as in the original array.

We begin with a discrete-time power spectral density $S_X(n)$ which
is periodic with period $M$, and with $S_X(0) = 0$.
For simulation purposes we define a conjugate symmetric
sequence $\tilde X_n$ based on the square root of $S_X(n)$:
\begin{equation}
\tilde X_n =
\left\{ \begin{array}{ll}
0 & \mbox{ for $n = 0$}\\
\exp(j \theta_n) \sqrt{S_X(n)} & \mbox{ for $0 < n < M/2$}\\
\sgn(\theta_n - \pi)\sqrt{S_X(n)} & \mbox{ for $n = M/2$}\\
\tilde X^*_{M-n} & \mbox{ for $M/2 < n < M$,}
\end{array} \right.
\end{equation}
where $\{\theta_n \}$ are independent and uniformly distributed in
$[0, 2\pi)$, $\sgn(\cdot)$ represents the signum function, and ${}^*$
denotes complex conjugation.
The corresponding time-domain signal may be obtained by inverse
Fourier transformation:
\begin{equation}
X_k \equiv M^{-1} \sum_{n=0}^{M-1} e^{j2\pi nk/M} \tilde X_n.
\end{equation}
We next define a subsampled sequence $Z_k$, also of length $M$, by
\begin{equation}
Z_k = \left\{ \begin{array}{ll}
X_{k/2} & \mbox{ for $k$ even}\\
X_{(k-1)/2} & \mbox{ for $k$ odd,}
\end{array} \right.
\end{equation}
for which the corresponding sequence in the frequency domain becomes
\begin{eqnarray}
\tilde Z_n & \equiv & \sum_{k=0}^{M-1}
Z_k e^{-j2\pi kn/M} \nonumber\\
& = & \sum_{l=0}^{M/2-1} X_l
\left[e^{-j2\pi (2l)n/M} + e^{-j2\pi (2l + 1)n/M} \right] \nonumber\\
& = & 2 e^{-j\pi n/M} \cos(\pi n/M)
\sum_{l=0}^{M/2-1} X_l e^{-j4\pi ln/M} \nonumber\\
& = & 2 M^{-1} e^{-j\pi n/M} \cos(\pi n/M)
\sum_{l=0}^{M/2-1} e^{-j4\pi ln/M}
\sum_{q=1}^{M-1} e^{j2\pi ql/M} \tilde X_q.
\end{eqnarray}
Finally, we compute the expected value of the periodogram of $Z$
\begin{eqnarray}
\E\left[\widehat{S_Z}(n)\right]
& \equiv & M^{-1} \E \left[\tilde Z_n \right]^2 \nonumber\\
& = & 4 M^{-3} \cos^2(\pi n/M)
\sum_{l=0}^{M/2-1} e^{-j4\pi ln/M} \sum_{k=0}^{M/2-1} e^{j4\pi kn/M}
\nonumber\\
&& \qquad \times
\sum_{q=1}^{M-1} e^{j2\pi ql/M} \sum_{r=1}^{M-1} e^{-j2\pi rk/M}
\E\left[\tilde X_q \tilde X^*_r \right] \nonumber\\
& = & 4 M^{-2} \cos^2(\pi n/M)
\sum_{l=0}^{M/2-1} e^{-j4\pi ln/M} \sum_{k=0}^{M/2-1} e^{j4\pi kn/M}
\sum_{q=1}^{M-1} e^{j2\pi ql/M} e^{-j2\pi qk/M}
S_X(q) \nonumber\\
& = & 4 M^{-2} \cos^2(\pi n/M) \sum_{q=1}^{M-1} S_X(q)
\left| \sum_{l=0}^{M/2-1} e^{-j2\pi (2n-q)l/M}\right|^2 \nonumber\\
& = & \cos^2(\pi n/M) \left\{S_X(2n) + 4 M^{-2}
\sum_{\stackrel{\scriptstyle q=1}
{q \mbox{\scriptsize \hspace{2pt} odd}}}^{M-1}
\csc^2\left[\pi(2n-q)/M\right]
S_X(q) \right\} .
\label{app_psd_sum}
\end{eqnarray}
Numerical computation based on Eq.~(\ref{app_psd_sum}) yields results
substantially in agreement with those of Eq.~(\ref{fractalpgex}).

\pagebreak

\section*{Figure Captions}

\begin{figure}[hb]
\vspace{1mm}
\caption{
Representations of a point process.
(a) The events are represented by a sequence of idealized impulses,
occurring at times $t_n$, and forming a stochastic point process
$dN(t)$.
For convenience of analysis, several alternative representations of
the point process are used.
(b) The counting process $N(t)$.
At every event occurrence the value of $N(t)$ augments by unity.
(c) The sequence of counts $\{ Z_{k} \}$, a discrete-time
non-negative integer-valued stochastic process, is formed from the
point process by recording the number of events in successive
counting windows of length $T$.
(d) The sequence of counts $\{ Z_{k} \}$ can be conveniently
described in terms of a count index $k$.
Information is lost because the precise times of event occurrences
within each counting window are eliminated in this representation.
Correlations in the discrete-time sequence $\{ Z_{k} \}$ can be
readily interpreted in terms of real time.
}
\label{basics}
%\end{figure}
% Keep as one big figure, to put on same page as "Figure Captions" title
%\begin{figure}
\vspace{1mm}
\caption{
Doubly logarithmic plot of the count-based periodogram vs.\ frequency
for the point process representing a spontaneous nerve spike train
recorded at the output of a visual-system relay neuron in the lateral
geniculate nucleus (LGN) of the cat (solid curve, cell No.\ MDS-LGN,
X-ON type {\bf [63]}).
%\cite{TEI96B}
No external stimulus was present.
The data segment analyzed here consists of 24285 events with an
average interevent time of 0.133 sec, comprising a total duration of
3225 sec.
Over long time scales (low frequencies), the curve can be fit by a
straight line (dotted) of slope -0.67, representing fractal behavior.
This least-squares straight-line fit to the data was calculated over
the region from 0.002 Hz to 1 Hz.
}
\label{vispp}
%\end{figure}
% Keep as one big figure, to put on same page as "Figure Captions" title
%\begin{figure}
\vspace{1mm}
\caption{
Doubly logarithmic plot of the Fano factor estimate vs.\ counting
time, for the same spike train as used in Fig.~2 (solid curve).
Over long time scales, the curve can be fit by a straight line
(dotted) of slope 0.66, representing fractal behavior with a similar
exponent to that obtained from the PG.
This straight-line best fit to the data (on doubly logarithmic
coordinates) was calculated over the region from 0.3 sec to 322 sec.
}
\label{visrff}
\end{figure}

\begin{figure}
\vspace{1mm}
\caption{
Doubly logarithmic plot of the Allan factor estimate vs.\ counting
time for the same spike train as used in Figs.~2 and~3 (solid curve).
Over long time scales, the curve can be fit by a straight line
(dotted) of slope 0.65, representing fractal behavior with a similar
exponent to that obtained from the PG and the FF.
This straight-line best fit to the data (on doubly logarithmic
coordinates) was calculated over the region from 0.3 sec to 322 sec.
}
\label{visaff}
\end{figure}

\begin{figure}
\vspace{1mm}
\caption{
Sample functions of fractal renewal processes.
Interevent times are power-law distributed.
(a) The standard fractal renewal process (FRP) consists of Dirac
$\delta$ functions and is zero-valued elsewhere.
(b) The alternating FRP switches between values of zero and unity.
}
\label{frp_diag}
\end{figure}

\begin{figure}
\vspace{1mm}
\caption{
A sum of several independent and identical
alternating FRPs (top) are added (center) to produce a fractal
binomial noise process which serves as the rate function for an
integrate-and-fire point process (bottom).
The result is the fractal-binomial-noise-driven integrate-and-fire
point process (FBNIF).
}
\label{fbndp_dg}
\end{figure}

\begin{figure}
\vspace{1mm}
\caption{
A primary homogeneous Poisson point process $M(t)$ with constant rate
$\mu$ serves as the input to a linear filter with impulse response
function $h(t)$.
The continuous-time stochastic process $I(t)$ at the output of this
filter is shot noise, which serves as the random rate for an
integrate-and-fire process whose output is $N(t)$.
$N(t)$ is the shot-noise driven integrate-and-fire point process
(SNIF).
If $h(t)$ decays in a power-law fashion, then $I(t)$ is fractal shot
noise and $N(t)$ is a fractal SNIF or FSNIF.
}
\label{fsndp_dg}
\end{figure}

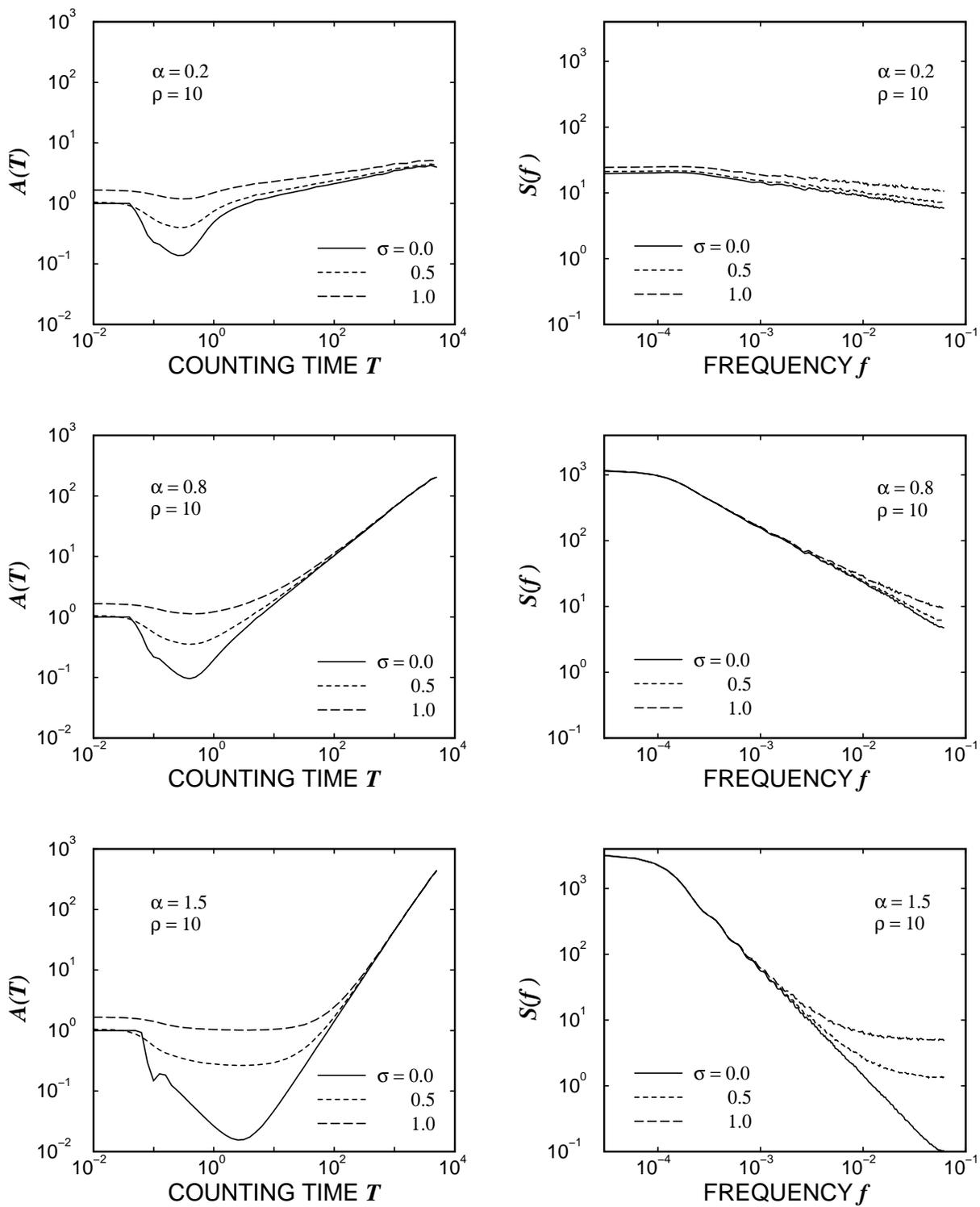
\begin{figure}
\vspace{1mm}
\caption{
AF curves (left) and PG curves (right) of the fractal-Gaussian-noise
driven integrate-and-fire process with jitter (FGNJIF), with
$\sigma=0$, $0.5$, and $1$ for three design values of the fractal
exponent: $\alpha=0.2$, $0.8$, and $1.5$.
The mean rate $\rho=10$ in all cases.
The underlying rate function $x[n]$ consisted of $2^{15}$ samples,
equally spaced at 1 sec each.
For $\alpha=0.2$ and $\alpha=0.8$ the FF intercept time $T_0$ was 1
sec; for $\alpha=1.5$ the PSD corner frequency $\omega_0$ was $0.005$
Hz.
Results from 100 simulations were averaged.
The jitter reduces the range of scaling and systematically lowers
the average slopes of the curves.
}
\label{gausse}
\end{figure}

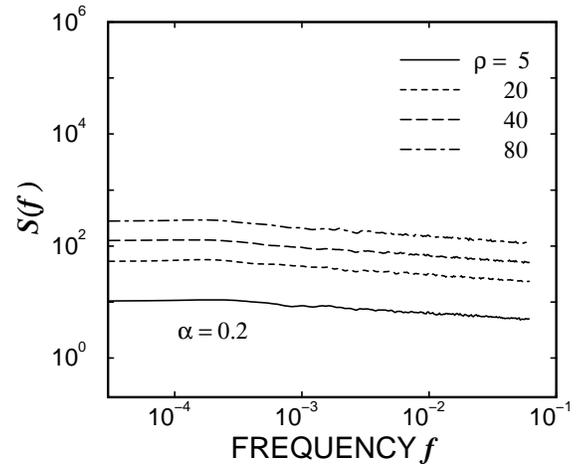
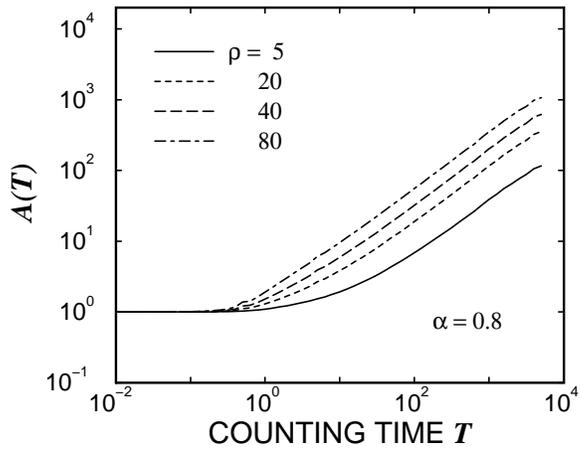
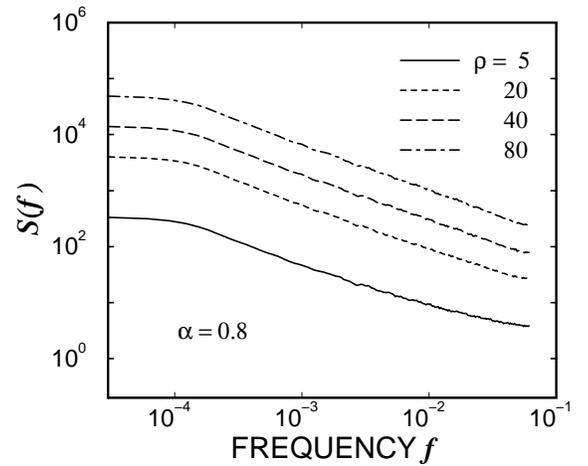
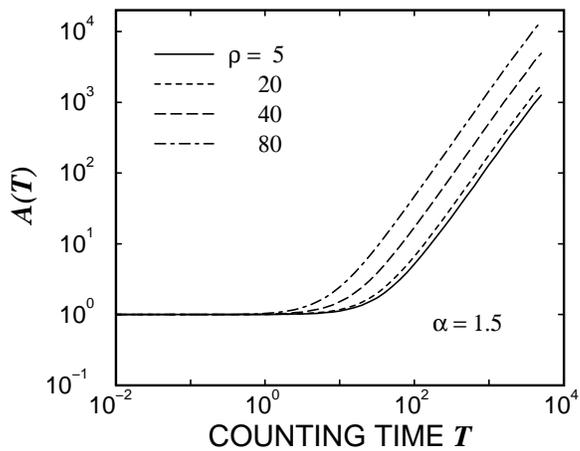
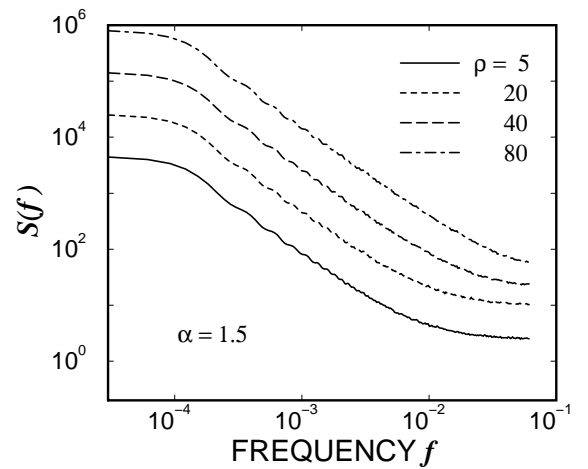
\begin{figure}
\vspace{1mm}
\caption{
AF curves (left) and PG curves (right) of the fractal-Gaussian-noise
driven Poisson process (FGNDP) with rate $\rho=5$, $20$, $40$, and
$80$, for three design values of the fractal exponent: $\alpha=0.2$,
$0.8$, and $1.5$.
The underlying rate function $x[n]$ consisted of $2^{15}$ samples,
equally spaced at 1 sec each.
For $\alpha=0.2$ and $\alpha=0.8$ the FF intercept time $T_0$ was 10
times the average interevent time $1/\rho$; for $\alpha=1.5$ the PSD
corner frequency $\omega_0$ was $0.0005$ times the average rate
$\rho$.
Results from 100 simulations were averaged.
As the rate decreases, the bias of the PG- and AF-based fractal
exponent estimators increases.
}
\label{poissons}
\end{figure}

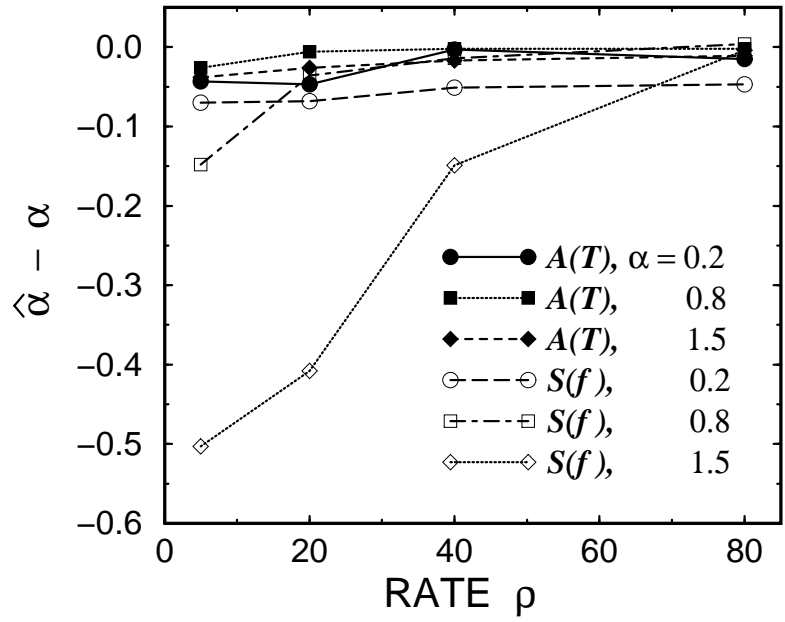
\begin{figure}
\vspace{1mm}
\caption{
Average differences between the estimated values of the fractal
exponents $\widehat{\alpha}$ and the design values $\alpha=0.2$,
$0.8$, and~$1.5$ as a function of the mean event production rate
$\rho$ for the fractal-Gaussian-noise driven Poisson process (FGNDP).
The AF-based estimate yields smaller bias than the PG-based measure.
For both fractal exponent estimators, bias decreases with increasing
average rate $\rho$, especially for the PG-based estimator with
$\alpha=1.5$.
}
\label{fehler}
\end{figure}


\begin{thebibliography}{99}
%
\bibitem{COX80} D.~R.~Cox and V.~Isham,
{\it Point Processes} (Chapman and Hall, London, 1980).
%
\bibitem{LOW95}
S.~B.~Lowen and M.~C.~Teich,
``Estimation and simulation of fractal stochastic point processes,''
{\it Fractals} {\bf 3}, 183-210 (1995).
%
\bibitem{RUD76} W.~Rudin,
{\it Principles of Mathematical Analysis,}
3rd ed.\ (McGraw Hill, New York, 1976), p.\ 197.
%
\bibitem{SHLE91}
M.~F.~Shlesinger and B.~J.~West,
``Complex fractal dimension of the bronchial tree,''
{\it Phys. Rev. Lett.} {\bf 67}, 2106-2108 (1991).
%
\bibitem{MAN83} B.~B.~Mandelbrot, {\it The Fractal Geometry of Nature}
(W.~H.~Freeman, New York, 1983).
%
\bibitem{HUR51} H.~E.~Hurst, ``Long term storage capacity of
reservoirs,'' {\it Trans.\ Amer.\ Soc.\ Civil Eng.} {\bf 116}, 770--808
(1951).
%
\bibitem{FLA89} P.~Flandrin,
``On the spectrum of fractional Brownian motions,''
{\it IEEE Trans.\ Inform.\ Theory} {\bf 35}, 197--199 (1989).
%
\bibitem{FLA92A} P.~Flandrin, ``Wavelet analysis and synthesis of
fractional Brownian motion,'' {\it IEEE Trans.\ Inform.\ Theory}
{\bf 38}, 910--917 (1992).
%
\bibitem{FLA92B} P.~Flandrin, ``Time--scale analyses and self-similar
stochastic processes,'' {\it Proc.\ NATO Advanced Study Institute on
Wavelets and their Applications} (Il Ciocco, Italy, 1992).
%
\bibitem{WOR92} G.~W.~Wornell and A.~V.~Oppenheim,
``Estimation of fractal signals from noisy measurements using
wavelets,''
{\it IEEE Trans.\ Sig.\ Proc.} {\bf 40}, 611--623 (1992).
%
\bibitem{BAR88} R.~J.~Barton and H.~V.~Poor, ``Signal detection in
fractional Gaussian noise,'' {\it IEEE Trans.\ Inform.\ Theory}
{\bf 34}, 943--959 (1988).
%
\bibitem{PRE78} W.~H.~Press,
``Flicker noises in astronomy and elsewhere,''
{\it Comm.\ Astrophys.} {\bf 7} (No.\ 4), 103--119 (1978).
%
\bibitem{BUC83} M.~J.~Buckingham,
{\it Noise in Electronic Devices and Systems}
(Wiley-Halsted, New York, 1983), Ch.\ 6.
%
\bibitem{WEI88} M.~B.~Weissman,
``$1/f$ noise and other slow, nonexponential kinetics in condensed
matter,'' {\it Rev.\ Mod.\ Phys.} {\bf 60}, 537--571 (1988).
%
\bibitem{BAS94}
J.~B.~Bassingthwaighte, L.~S.~Liebovitch, and B.~J.~West,
{\it Fractal Physiology}
(American Physioloical Society, New York, 1994).
%
\bibitem{WES94} B.~J.~West and W.~Deering,
``Fractal physiology for physicists: L\'evy statistics,''
{\it Phys.\ Rep.} {\bf 246}, 1--100 (1994).
%
\bibitem{AND96} C.~M.~Anderson and A.~J.~Mandell,
``Fractal time and the foundations of consciousness: vertical
convergence of $1/f$ phenomena from ion channels to behavioral
states,''
in {\it Fractals of Brain, Fractals of Mind}
(Advances in Consciousness Research {\bf 7}),
eds.\ E.~MacCormac and M.~Stamenov
(John Benjamin, Amsterdam, 1996), pp.\ 75--128.
%
\bibitem{PFI78} G.~Pfister and H.~Scher,
``Dispersive (non-Gaussian) transient transport in disordered
solids,''
{\it Adv.\ Phys.} {\bf 27}, 747--798 (1978).
%
\bibitem{ORE82} J.~Orenstein, M.~A.~Kastner, and V.~Vaninov,
``Transient photoconductivity and photo-induced optical absorption in
amorphous semiconductors,''
{\it Phil.\ Mag.\ B} {\bf 46}, 23--62 (1982).
%
\bibitem{KAS85} M.~A.~Kastner,
``The peculiar motion of electrons in amorphous semiconductors,''
in {\it Physical Properties of Amorphous Materials,}
eds. D.~Alser, B.~B.~Schwartz, and M.~C.~Steele,
(Plenum, New York, 1985), pp.\ 381--396.
%
\bibitem{TIE80} T.~Tiedje and A.~Rose,
``A physical interpretation of dispersive transport in disordered
semiconductors,''
{\it Solid State Commun.} {\bf 37}, 49--52 (1980).
%
\bibitem{SHL87} M.~F.~Shlesinger,
``Fractal time and $1/f$ noise in complex systems,''
{\it Ann.\ New York Acad.\ Sci.} {\bf 504}, 214--228 (1987).
%
\bibitem{LOW93A} S.~B.~Lowen and M.~C.~Teich, ``Fractal renewal
processes generate $1/f$ noise,''
{\it Phys.\ Rev.\ E} {\bf 47}, 992--1001 (1993).
%
\bibitem{LOW92A} S.~B.~Lowen and M.~C.~Teich,
``Fractal renewal processes as a model of charge transport in
amorphous semiconductors,''
{\it Phys.\ Rev.\ B} {\bf 46}, 1816--1819 (1992).
%
\bibitem{BHA90} V.~K.~Bhatnagar and K.~L.~Bhatia,
``Frequency dependent electrical transport in bismuth-modified
amorphous germanium sulfide semiconductors,''
{\it J. Non-cryst.\ Sol.} {\bf 119}, 214--231 (1990).
%
\bibitem{TOM90} W.~Tomaszewicz,
``Multiple-trapping carrier transport due to modulated
photogeneration,''
{\it Phil.\ Mag.\ Lett.} {\bf 61}, 237--243 (1990).
%
\bibitem{BER63} J.~M.~Berger and B.~B.~Mandelbrot,
``A new model for the clustering of errors on telephone circuits,''
{\it IBM J. Res.\ Dev.} {\bf 7}, 224--236 (1963).
%
\bibitem{MAN65} B.~B.~Mandelbrot,
``Self-similar error clusters in communication systems and the
concept of conditional stationarity,''
{\it IEEE Trans.\ Comm.\ Tech.} {\bf 13}, 71--90 (1965).
%
\bibitem{LEL93} W.~E.~Leland, M.~S.~Taqqu, W.~Willinger, and
D.~V.~Wilson,
``On the self-similar nature of Ethernet traffic,'' in
{\it Proc.\ ACM SIGCOMM 1993}, (1993), pp.\ 183--193.
%
\bibitem{ERR93} A.~Erramilli and W.~Willinger,
``Fractal properties in packet traffic measurements,'' in
{\it Proc.\ St.\ Petersburg Reg.\ ITC Sem.},
(St.\ Petersburg, Russia, 1993), pp.\ 144-158.
%
\bibitem{RYU94} B.~K.~Ryu and H.~E.~Meadows,
``Performance analysis and traffic behavior of Xphone
videoconferencing application on an Ethernet,'' in
{\it Proc.\ 3rd Int.\ Conf.\ Comp.\ Commun.\ Netw.},
ed.\ W.~Liu, (1994), pp.\ 321--326.
%
\bibitem{LEL94} W.~E.~Leland, M.~S.~Taqqu, W.~Willinger, and
D.~V.~Wilson, ``On the self-similar nature of Ethernet traffic
(extended version),''
{\it IEEE/ACM Trans.\ Netw.} {\bf 2}, 1--15 (1994).
%
\bibitem{BER95} J.~Beran, R.~Sherman, M.~S.~Taqqu, and W.~Willinger,
``Long-range dependence in variable-bit-rate video traffic,''
{\it IEEE Trans.\ Comm.} {\bf 43}, 1566--1579 (1995).
%
\bibitem{RYU95} B.~K.~Ryu and S.~B.~Lowen,
``Modeling self-similar traffic with the fractal-shot-noise-driven
Poisson process,'' Cent.\ for Telecomm.\ Res., Tech.\ Rep.\
392-94-39, (Columbia University, New York, 1994).
%
\bibitem{SAK83} B.~Sakmann and E.~Neher,
{\it Single-Channel Recording} (Plenum, New York, 1983).
%
\bibitem{DEF93} L.~J.~DeFelice and A.~Isaac,
``Chaotic states in a random world: relationship between the nonlinear
differential equations of excitability and the stochastic properties
of ion channels,''
{\it J. Stat.\ Phys.} {\bf 70}, 339--354 (1993).
%
\bibitem{LAU88} P.~L\"auger,
``Internal motions in proteins and gating kinetics of ionic
channels,''
{\it Biophys.\ J.} {\bf 53}, 877--884 (1988).
%
\bibitem{MIL88} G.~L.~Millhauser, E.~E.~Salpeter, and R.~E.~Oswald,
``Diffusion models of ion-channel gating and the origin of power-law
distributions from single-channel recording,''
{\it Proc.\ Natl.\ Acad.\ Sci.} (USA) {\bf 85}, 1503--1507 (1988).
%
\bibitem{LIE90} L.~S.~Liebovitch and T.~I.~T\'{o}th,
``Using fractals to understand the opening and closing of ion
channels,''
{\it Ann.\ Biomed.\ Eng.} {\bf 18}, 177--194 (1990).
%
\bibitem{TEI89} M.~C.~Teich,
``Fractal character of the auditory neural spike train,''
{\it IEEE Trans.\ Biomed.\ Eng.} {\bf 36}, 150--160 (1989).
%
\bibitem{LOW93B} S.~B.~Lowen and M.~C.~Teich, ``Fractal renewal
processes,''
{\it IEEE Trans.\ Inform.\ Theory} {\bf 39}, 1669--1671 (1993).
%
\bibitem{LOW93C} S.~B.~Lowen and M.~C.~Teich, ``Fractal
auditory-nerve firing patterns may derive from fractal switching
in sensory hair-cell ion channels,''
in {\it Noise in Physical Systems and $1/f$ Fluctuations}
(AIP Conference Proceedings {\bf 285}),
eds.\ P.~H.~Handel and A.~L.~Chung,
(American Institute of Physics, New York, 1993), pp.\ 781--784.
%
\bibitem{KAT66} B.~Katz, {\it Nerve, Muscle, and Synapse}
(McGraw-Hill, New York, 1966).
%
\bibitem{FAT52} P.~Fatt and B.~Katz,
``Spontaneous subthreshold activity at motor nerve endings,''
{\it J. Physiol.} (London) {\bf 117}, 109--128 (1952).
%
\bibitem{DEL54} J.~Del~Castillo and B.~Katz,
``Quantal components of the end-plate potential,''
{\it J. Physiol.} (London) {\bf 124}, 560--573 (1954).
%
\bibitem{LOW97A} S.~B.~Lowen, S.~S.~Cash, M-m.~Poo, and M.~C.~Teich,
``Neuronal Exocytosis Exhibits Fractal Behavior,''
in {\it Computational Neuroscience,} ed.\ J.~M.~Bower
(Plenum, New York, 1997), in press.
%
\bibitem{LOW97B} S.~B.~Lowen, S.~S.~Cash, M-m.~Poo, and M.~C.~Teich,
``Quantal Neurotransmitter Secretion Rate Exhibits Fractal Behavior,"
{\it J. Neurosci.}, in press.
%
%
\bibitem{MUS81} T.~Musha, Y.~Kosugi, G.~Matsumoto, and M.~Suzuki,
``Modulation of the time relation of action potential impulses
propagating along an axon,''
{\it IEEE Trans.\ Biomed.\ Eng.} {\bf BME-28}, 616--623 (1981).
%
\bibitem{MUS83} T.~Musha, H.~Takeuchi, and T.~Inoue,
``$1/f$ fluctuations in the spontaneous spike discharge intervals of
a giant snail neuron,''
{\it IEEE Trans.\ Biomed.\ Eng.} {\bf BME-30}, 194--197 (1983).
%
%
\bibitem{LOW92B} S.~B.~Lowen and M.~C.~Teich,
``Auditory-nerve action potentials form a non-renewal point process
over short as well as long time scales,''
{\it J.\ Acoust.\ Soc.\ Am.\ } {\bf 92}, 803--806 (1992).
%
\bibitem{TEI90B}
M.~C.~Teich, D.~H.~Johnson, A.~R.~Kumar, and R.~G.~Turcott,
``Rate fluctuations and fractional power-law noise recorded from
cells in the lower auditory pathway of the cat,''
{\it Hear.\ Res.} {\bf 46}, 41--52 (1990).
%
\bibitem{TEI90A} M.~C.~Teich, R.~G.~Turcott, and S.~B.~Lowen,
``The fractal doubly stochastic Poisson point process as a model for
the cochlear neural spike train,''
in {\it The Mechanics and Biophysics of Hearing (Lecture Notes in
Biomathematics, Vol.\ 87),} eds.\ P.~Dallos,
C.~D.~Geisler, J.~W.~Matthews, M.~A.~Ruggero, and C.~R.~Steele
(Springer-Verlag, New York, 1990), pp.\ 354--361.
%
\bibitem{POW91}
N.~L.~Powers, R.~J.~Salvi, and S.~S.~Saunders,
``Discharge rate fluctuations in the auditory nerve of the
chinchilla,''
in {\it Abstracts of the Fourteenth Midwinter Research Meeting of the
Association for Research in Otolaryngology,} ed.\ D.~J.~Lim
(Association for Research in Otolaryngology, Des Moines, IA),
Abstract 411, p.\ 129.
%
\bibitem{POW92}
N.~L.~Powers and R.~J.~Salvi,
``Comparison of discharge rate fluctuations in the auditory nerve of
chickens and chinchillas,''
in {\it Abstracts of the Fifteenth Midwinter Research Meeting of the
Association for Research in Otolaryngology,} ed.\ D.~J.~Lim
(Association for Research in Otolaryngology, Des Moines, IA, 1992),
Abstract 292, p.\ 101.
%
\bibitem{TEI92} M.~C.~Teich, ``Fractal neuronal firing patterns,''
in {\it Single Neuron Computation,} eds.\ T.~McKenna, J.~Davis, and
S.~Zornetzer (Academic, Boston, 1992), pp.\ 589--625.
%
\bibitem{KUM93} A.~H.~Kumar and D.~H.~Johnson,
``Analyzing and modeling fractal intensity point processes,''
{\it J. Acoust.\ Soc.\ Am.} {\bf 93}, 3365--3373 (1993).
%
\bibitem{TEI94} M.~C.~Teich and S.~B.~Lowen,
``Fractal patterns in auditory nerve-spike trains,''
{\it IEEE Eng.\ Med.\ Biol.\ Mag.} {\bf 13} (No.\ 2), 197--202 (1994).
%
\bibitem{KEL94}
O.~E.~Kelly,
``Analysis of long-range dependence in auditory-nerve fiber
recordings,''
Master's Thesis, Rice University (1994).
%
\bibitem{KEL96}
O.~E.~Kelly, D.~H.~Johnson, B.~Delgutte, and P.~Cariani,
``Fractal noise strength in auditory-nerve fiber recordings,''
{\it J. Acoust.\ Soc.\ Am.} {\bf 99}, 2210--2220 (1996).
%
\bibitem{LOW96B} S.~B.~Lowen and M.~C.~Teich,
``The periodogram and Allan variance reveal fractal exponents
greater than unity in auditory-nerve spike trains,''
{\it J. Acoust.\ Soc.\ Am} {\bf 99}, 3585--3591 (1996).
%
\bibitem{LOW96D} S.~B.~Lowen and M.~C.~Teich,
``Refractoriness-modified fractal stochastic point processes for
modeling sensory-system spike trains,''
in {\it Computational Neuroscience,} ed.\ J.~M.~Bower
(Academic, New York, 1996), pp.\ 447-452.
%
\bibitem{LOW96A} S.~B.~Lowen and M.~C.~Teich,
``Estimating scaling exponents in auditory-nerve spike trains
using fractal models incorporating refractoriness,''
in {\it Diversity in Auditory Mechanics,}
eds.\ E.~R~Lewis, G.~R.~Long, R.~F.~Lyon, P.~M.~Narins, C.~R.~Steele,
and E. Hecht-Pointar (World Scientific, Singapore, 1997),
pp. 197-204.
%
\bibitem{TEI96B} M.~C.~Teich, C.~Heneghan, S.~B.~Lowen, T.~Ozaki,
and E.~Kaplan, ``Fractal character of the neural spike train in the
visual system of the cat,''
{\it J. Opt.\ Soc.\ Am.\ A} {\bf 14}, 529--546 (1997).
%
\bibitem{TRO92} J.~B.~Troy and J.~G.~Robson,
``Steady discharges of X and Y retinal ganglion cells of cat under
photopic illuminance,''
{\it Visual Neuroscience} {\bf 9}, 535--553 (1992).
%
\bibitem{TEI95} M.~C.~Teich, R.~G.~Turcott, and R.~M.~Siegel,
``Temporal correlation in cat striate-cortex neural spike trains,''
{\it IEEE Eng.\ Med.\ Biol.\ Mag.} {\bf 15} (No.\ 5), 79--87 (1996).
%
\bibitem{TUR95} R.~G.~Turcott, P.~D.~R.~Barker, and M.~C.~Teich,
``Long-duration correlation in the sequence of action potentials in
an insect visual interneuron,'' {\it J. Statist.\ Comput.\ Simul.}
{\bf 52}, 253--271 (1995).
%
\bibitem{WIS81} M.~E.~Wise,
``Spike interval distributions for neurons and random walks with
drift to a fluctuating threshold,''
in {\it Statistical Distributions in Scientific Work, Vol.\ 6},
eds.\ C.~Taillie, G.~P.~Patil, and B.~A.~Baldessari
(D.~Reidel, Hingham, MA, 1981), pp.\ 211--231.
%
\bibitem{YAM86}
M.~Yamamoto, H.~Nakahama, K.~Shima, T.~Kodama, and H.~Mushiake,
``Markov-dependency and spectral analyses on spike-counts in
mesencephalic reticular formation during sleep and attentive states,''
{\it Brain Research} {\bf 366}, 279--289 (1986).
%
\bibitem{GRU89} F.~Gr\"uneis, M.~Nakao, M.~Yamamoto, T.~Musha, and
H.~Nakahama,
``An interpretation of $1/f$ fluctuations in neuronal spike trains
during dream sleep,''
{\it Biol.\ Cybern} {\bf 60}, 161--169 (1989)
%
\bibitem{GRU93} F.~Gr\"uneis, M.~Nakao, Y.~Mizutani, M.~Yamamoto,
M.~Meesman, and T.~Musha,
``Further study on $1/f$ fluctuations observed in central single
neurons during REM sleep,''
{\it Biol.\ Cybern.} {\bf 68} 193--198 (1993).
%
\bibitem{KOD89}
T.~Kodama, H.~Mushiake, K.~Shima, H.~Nakahama, and M.~Yamamoto,
``Slow fluctuations of single unit activities of hippocampal and
thalamic neurons in cats. I. Relation to natural sleep and alert
states,''
{\it Brain Research} {\bf 487}, 26--34 (1989).
%
\bibitem{YAM83} M.~Yamamoto and H.~Nakahama,
``Stochastic properties of spontaneous unit discharges in
somatosensory cortex and mesencephalic reticular formation during
sleep--waking states,''
{\it J. Neurophysiology} {\bf 49}, 1182--1198 (1983).
%
\bibitem{KOB82} M.~Kobayashi and T.~Musha,
``$1/f$ fluctuations of heartbeat period,''
{\it IEEE Trans.\ Biomed.\ Eng.} {\bf BME-29}, 456--457 (1982).
%
\bibitem{BER86} R.~D.~Berger, S.~Akselrod, D.~Gordon, and R.~J.~Cohen,
``An efficient algorithm for spectral analysis of heart rate
variability,''
{\it IEEE Trans.\ Biomed.\ Eng.} {\bf BME-33}, 900--904 (1986).
%
\bibitem{TUR93} R.~G.~Turcott and M.~C.~Teich,
``Long-duration correlation and attractor topology of the heartbeat
rate differ for healthy patients and those with heart failure,''
{\it Proc.\ SPIE} {\bf 2036} (Chaos in Biology and
Medicine), 22--39 (1993).
%
\bibitem{TUR96} R.~G.~Turcott and M.~C.~Teich,
``Interevent-interval and counting statistics of the human heartbeat
recorded from normal subjects and patients with heart failure,''
{\it Ann.\ Biomed.\ Eng.} {\bf 24}, 269--293 (1996).
%
\bibitem{SCH92} H.~E.~Schepers, J.~H.~G.~M.~van~Beek, and
J.~B.~Bassingthwaighte, ``Four methods to estimate the fractal dimension
from self-affine signals,'' {\it IEEE Eng.\ Med.\ Biol.\ Mag.}
{\bf 11}, 57--64 (1992).
%
\bibitem{BER92} J.~Beran,
``Statistical methods for data with long-range dependence,''
{\it Statistical Science} {\bf 7}, 404--427 (1992).
%
\bibitem{HEN83} H.~G.~E.~Hentschel and I.~Procaccia,
``The infinite number of generalized dimensions of fractals and
strange attractors,''
{\it Physica D} {\bf 8}, 435--444 (1983).
%
\bibitem{GRA83} P.~Grassberger,
``Generalized dimensions of strange attractors,''
{\it Phys.\ Lett.\ A} {\bf 97}, 227-230 (1983).
%
\bibitem{THE90} J.~Theiler,
``Estimating fractal dimension,''
{\it J. Opt.\ Soc.\ Am.\ A} {\bf 7}, 1055--1073 (1990).
%
\bibitem{MAT82} K.~Matsuo, B.~E.~A.~Saleh, and M.~C.~Teich,
``Cascaded Poisson processes,''
{\it J. Math.\ Phys.} {\bf 23}, 2353--2364 (1982).
%
\bibitem{LOW93D} S.~B.~Lowen and M.~C.~Teich,
``Estimating the dimension of a fractal point processes,''
{\it Proc.\ SPIE} {\bf 2036} (Chaos in Biology and
Medicine), 64--76 (1993).
%
\bibitem{ALL66} D.~W.~Allan, ``Statistics of atomic frequency
standards,'' {\it Proc.\ IEEE}$\;$ {\bf 54}, 221-230 (1966).
%
\bibitem{BAR66} J.~A.~Barnes and D.~W.~Allan, ``A statistical model
of flicker noise,'' {\it Proc.\ IEEE} {\bf 54}, 176--178 (1966).
%
\bibitem{TEI96A}
M.~C.~Teich, C.~Heneghan, S.~B.~Lowen, and R.~G.~Turcott,
``Estimating the fractal exponent of point processes in biological
systems using wavelet- and Fourier-transform methods,''
in {\it Wavelets in Medicine and Biology}, eds.\ A.~Aldroubi and
M.~Unser (CRC Press, Boca Raton, FL, 1996), pp.\ 383--412.
%
\bibitem{HEN96} C.~Heneghan, S.~B.~Lowen, and M.~C.~Teich,
``Wavelet analysis for estimating the fractal properties of neural
firing patterns,''
in {\it Computational Neuroscience}, ed.\ J.~M.~Bower,
(Academic Press, San Diego, 1996), pp.\ 441--446.
%
\bibitem{LOW96C}
S.~B.~Lowen,
``Refractoriness-Modified Doubly Stochastic Poisson Point Process,''
{\it Center for Telecommunications Research, Technical Report}
{\bf 449-96-15} (Columbia University, New York, 1996).
%
\bibitem{LOW91} S.~B.~Lowen and M.~C.~Teich, ``Doubly stochastic
Poisson point process driven by fractal shot noise,''
{\it Phys.\ Rev.\ A} {\bf 43}, 4192--4215 (1991).
%
\bibitem{GRU86A} F.~Gr\"{u}neis and H.-J.~Baiter,
``More detailed explication of a number fluctuation model generating
$1/f$ pattern,'' {\it Physica} {\bf 136A}, 432--452 (1986).
%
\bibitem{HAI67} F.~A.~Haight,
{\it Handbook of the Poisson Distribution} (Wiley, New York, 1967).
%
\bibitem{MAN71} B.~B.~Mandelbrot,
``A fast fractional Gaussian noise generator,''
{\it Water Resources Res.} {\bf 7}, 543--553 (1971).
%
\bibitem{LUN86} T.~Lundahl, W.~J.~Ohley, S.~M.~Kay, and R.~Siffert,
``Fractional Brownian motion: a maximum likelihood estimator and
its application to image texture,''
{\it IEEE Trans.\ Med.\ Imag.} {\bf 5}, 152--161 (1986).
%
\bibitem{STO94} M.~A.~Stoksik, R.~G.~Lane, and D.~T.~Nguyen,
``Accurate synthesis of fractional Brown\-ian motion using wavelets,''
{\it Electron.\ Lett.} {\bf 30}, 383--384 (1994).
%
\bibitem{SAL78} B.~E.~A.~Saleh, {\it Photoelectron Statistics}
(Springer Verlag, Berlin, 1978).
%
\bibitem{LOW89A} S.~B.~Lowen and M.~C.~Teich,
``Generalised $1/f$ shot noise,'' {\it Electron.\ Lett.}
{\bf 25}, 1072--1074 (1989).
%
\bibitem{LOW89B} S.~B.~Lowen and M.~C.~Teich,
``Fractal shot noise,'' {\it Phys.\ Rev.\ Lett.}
{\bf 63}, 1755--1759 (1989).
%
\bibitem{LOW90} S.~B.~Lowen and M.~C.~Teich,
``Power-law shot noise,'' {\it IEEE Trans.\ Inform.\ Theory}
{\bf 36}, 1302--1318 (1990).
%
\bibitem{COX55} D.~R.~Cox,
``Some statistical methods connected with series of events,''
{\it J. Roy.\ Stat.\ Soc.\ B} {\bf 17}, 129--164 (1955).
%
\bibitem{GRU84} F.~Gr\"{u}neis,
``A number fluctuation model generating $1/f$ pattern,''
{\it Physica} {\bf 123A}, 149--160 (1984).
%
\bibitem{GRU86B} F.~Gr\"{u}neis and T.~Musha,
``Clustering Poisson process and $1/f$ noise,''
{\it Jpn.\ J. Appl.\ Phys.} {\bf 25}, 1504--1509 (1986).
%
\bibitem{SKIL88}
J.~Skilling, ed.,
{\it Maximum Entropy and Bayesian Methods}
(Kluwer, Boston, 1988).
\end{thebibliography}
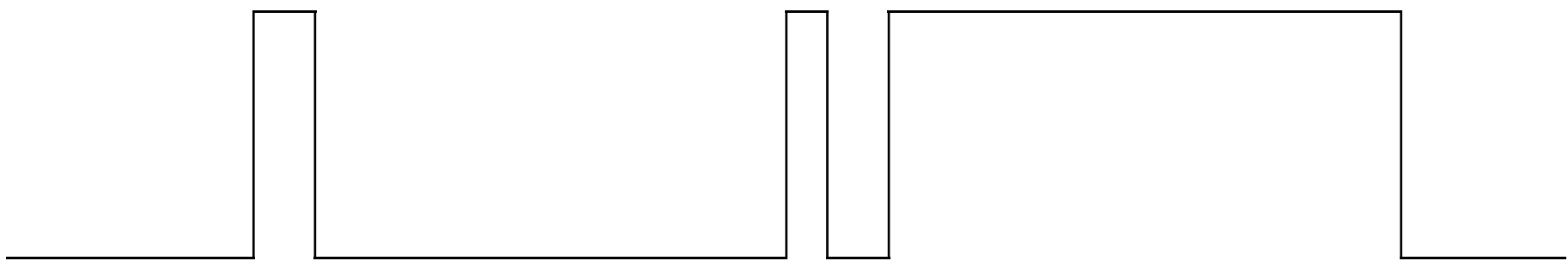
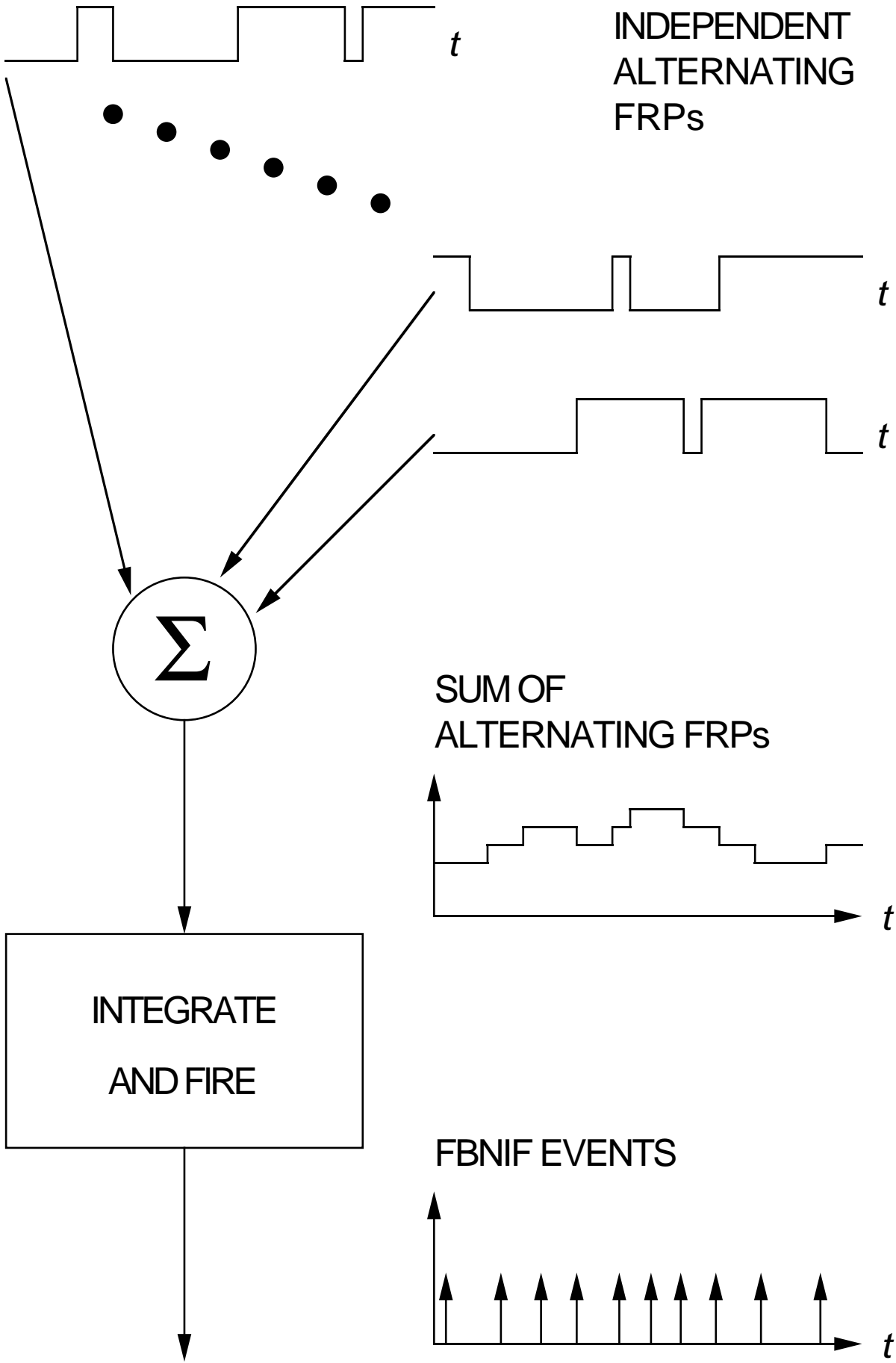
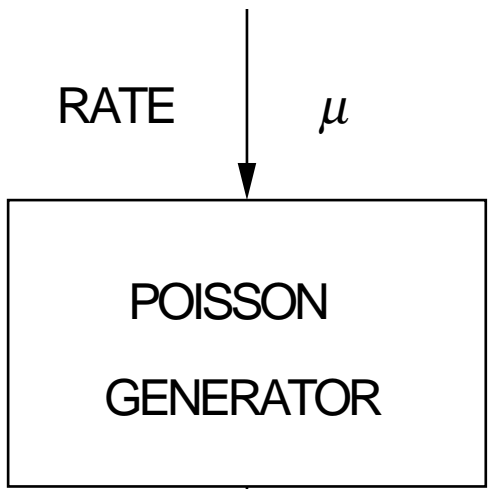
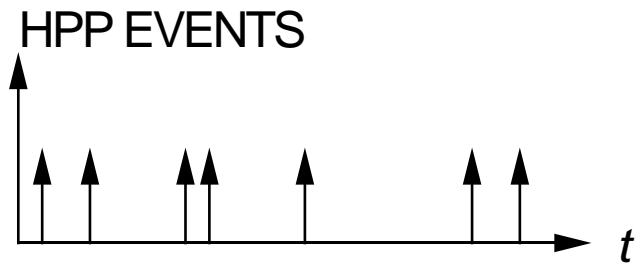
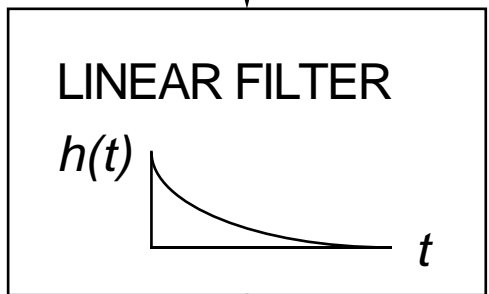
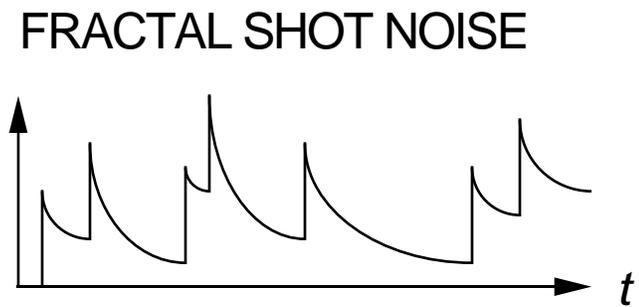
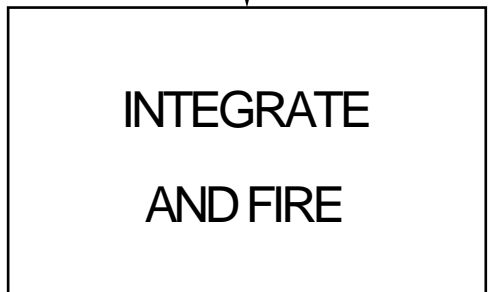
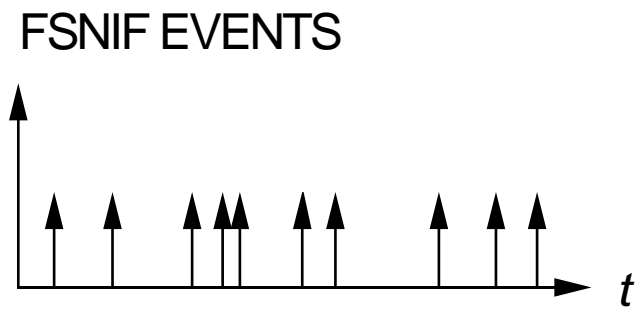
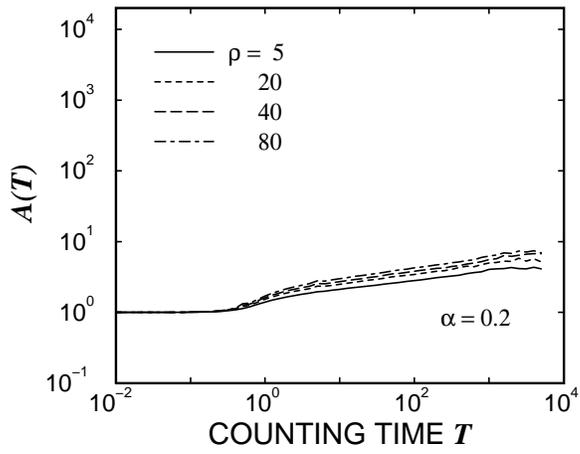
\end{document}